\documentclass[article]{JHEP3}
\usepackage{graphicx}
\usepackage{cite}
\usepackage{multirow, multicol}
\usepackage{hep}

\newcommand{\massboundsneutral}{CMS-PAS-HIG-13-002,ATLAS:2013nma,CMS-PAS-HIG-13-021,CMS-PAS-HIG-13-014,ATLAS:2013wla,Aad:2012cfr,CMS-PAS-HIG-13-025,Mankel:2014gwa,Aad:2014yja,CMS:2014ipa}
\newcommand{\massboundscharged}{Aad:2012tj,Chatrchyan:2012vca,ATLAS-CONF-2013-090,Abbiendi:2013hk}

\newcommand{\unitarity}{Akeroyd:2000wc,Kanemura:1993hm,Maalampi:1991fb,Ginzburg:2005dt,Osland:2008aw}
\newcommand{\flavor}{Aubert:2007my,Su:2009fz,Aoki:2009ha,Mahmoudi:2012ej}
\newcommand{\vacuum}{Gunion:2002zf,Kanemura:1999xf,Kastening:1992by,Velhinho:1994np,Nie:1998yn,Ferreira:2004yd,Maniatis:2006fs,Ferreira:2009jb}
\newcommand{\ewpo}{Toussaint:1978zm,Frere:1982ma,Bertolini:1985ia,Hollik:1986gg,Hollik:1987fg,Froggatt:1991qw,Grimus:2007if,Grimus:2008nb,He:2001tp}
\newcommand{\fits}{Ferreira:2011aa,Ferreira:2012my,Azatov:2012wq,Alves:2012ez,Cheon:2012rh,Bai:2012ex,Chiang:2013ixa,Grinstein:2013npa,Krawczyk:2013gia,Barroso:2013zxa,Coleppa:2013dya,Chen:2013rba,Chen:2013kt,Belanger:2013xza,Celis:2013rcs,Altmannshofer:2012ar,Craig:2013hca,Barger:2013ofa,Celis:2013ixa,Cheung:2013rva,Chang:2013ona,Wang:2013sha,Cheng:2014ova,Coleppa:2014hxa}
\newcommand{\collider}{Aoki:2011wd,Chang:2011kr,Arhrib:2011wc,Kanemura:2011kx,Mader:2012pm,Tsumura:2013yba,Harlander:2013mla,Harlander:2013qxa,Bagnaschi:2014zla,Kanemura:2014bqa}
\newcommand{\superiso}{Mahmoudi:2007vz,Mahmoudi:2008tp}
\newcommand{\higgsbounds}{Bechtle:2008jh,Bechtle:2011sb}

 \newcommand{\sda}{\ensuremath{\sin 2\alpha}}
 
  \newcommand{\sba}{\ensuremath{\sin(\beta-\alpha)}}
  \newcommand{\cba}{\ensuremath{\cos(\beta-\alpha)}}
  \newcommand{\sdb}{\ensuremath{\sin\,2\beta}}
   \newcommand{\sbad}{\ensuremath{\sin^2(\beta-\alpha)}}
  
 \newcommand{\sab}{\ensuremath{\sin(\alpha+\beta)}}
 \newcommand{\cab}{\ensuremath{\cos(\alpha+\beta)}}
\newcommand{\mhd}{\ensuremath{m^2_{\hzero}}}
\newcommand{\mHHd}{\ensuremath{m^2_{\Hzero}}}
\newcommand{\motd}{\ensuremath{m^2_{12}}}

\newcommand{\hzero}{\ensuremath{\PHiggslightzero}} 
\newcommand{\Hzero}{\ensuremath{\PHiggsheavyzero}} 
\newcommand{\Azero}{\ensuremath{\PHiggspszero}} 
\newcommand{\lag}{\ensuremath{\mathcal{L}}}

\newcommand{\CP}{\ensuremath{\mathcal{C}\mathcal{P}}}

\preprint{MCnet-14-12,CP3-14-40}
\title{Higgs pair production via gluon fusion in the Two--Higgs--Doublet Model}
\author{Beno\^it Hespel, David L\'opez-Val, Eleni Vryonidou \\Center for Cosmology, 
 Particle Physics and Phenomenology (CP3),\\
 Universit\'e catholique de Louvain, B-1348 Louvain-la-Neuve, Belgium\\

 Emails: \email{benoit.hespel@uclouvain.be}, \email{david.lopezval@uclouvain.be}, \email{eleni.vryonidou@uclouvain.be}}
\abstract
{We study the production of Higgs boson pairs via gluon fusion 
at the LHC in the Two--Higgs--Doublet Model.
We present predictions 
at NLO accuracy in QCD,
matched to parton showers through the MC$@$NLO method.
A dedicated
reweighting technique is used 
to improve the NLO calculation upon the infinite top--mass limit.
We perform our calculation within the MadGraph5$\_${ \sc aMC@NLO} framework, 
along with the {\sc 2HDM}
implementation based on the {\sc NLOCT} package. 
The inclusion of the NLO corrections leads to large $K$--factors 
and significantly reduced theoretical uncertainties.
We examine the seven 2HDM Higgs pair combinations
using a number of 
representative 2HDM scenarios. We show
how the model--specific features modify the Higgs pair total rates 
and distribution shapes, leading to trademark signatures of an extended Higgs sector.}

\maketitle

\begin{document}

\section{Introduction}
\label{sec:intro}

Recent LHC data provide evidence that the scalar particle
observed at the LHC is the one predicted by the electroweak symmetry breaking mechanism~\cite{Englert:1964et,Higgs:1964pj},
as implemented in the Standard Model (SM)~\cite{Higgs:1964pj}.
The mechanism predicts the strengths of all Higgs boson couplings to be uniquely
determined by the masses of the elementary particles. This also includes 
the triple and quartic Higgs boson self--couplings, which in the SM are linked to the Higgs
mass and the vacuum expectation value (VEV), and are 
therefore fully fixed after the Higgs mass measurement.
Current measurements of the Higgs couplings to fermions and vector bosons \cite{cms2013,atlas2013} agree within 10-20\%
with the SM predictions, while no direct information is
available on the Higgs boson self--interactions.
In view of this experimental picture, possible non--minimal Higgs sectors are
constrained, but certainly not ruled out. 
While some new physics models are no longer compatible with the
present collider data, there are still 
many extended Higgs sectors which can accommodate a $\sim 126$ GeV Higgs boson  
with coupling strengths similar to the SM.
One prime candidate for such a theory is the Two--Higgs--Doublet Model (2HDM)~\cite{Branco:2011iw}. 
This model provides a simple
UV--complete perturbative extension of the SM, which we can view as 
the low--energy Higgs sector of more fundamental
theories such as the 
Minimal Supersymmetric Standard Model (MSSM)~\cite{Carena:2002es,Gunion:1989we},
GUTs~\cite{Stech:2012zr,Stech:2014jda}, composite Higgs
models~\cite{Kaplan:1983sm,Agashe:2004rs,Burdman:2011fw}, and little Higgs models~\cite{Schmaltz:2005ky,Perelstein:2005ka}.
Aside from its very rich and distinctive phenomenology, 
the 2HDM also sets the ground for novel approaches
to diverse unsettled conundrums, from e.g. the origin of neutrino 
masses~\cite{Ma:2006km} to Naturalness ~\cite{Chakraborty:2014oma}, Electroweak
Baryogenesis~\cite{Kanemura:2004ch,Cline:2011mm,Tranberg:2012jp,Dorsch:2013wja} and Dark
Matter~\cite{LopezHonorez:2006gr,Honorez:2010re,Gong:2012ri}. 
If effectively realized in Nature, the 2HDM could
lead to genuine indications of new physics at colliders.
For instance, 
the direct production of additional heavy scalars
offers excellent prospects to identify these novel particles
through a variety of decay modes and final--state signatures \cite{\collider}.
Not less important are the indirect signatures from such extended Higgs sectors,
which could arise via modified Higgs couplings to the
fermions and gauge bosons, as well as through modified Higgs self--interactions.

\smallskip{}
At present, the experimental reconstruction of the Higgs potential
is a major step towards the fundamental understanding of
the EW symmetry breaking mechanism.  
This task can only be fully accomplished 
through the direct measurement of the Higgs boson three--point
and four--point interactions \cite{Djouadi:1999rca,Gupta:2013zza,Efrati:2014uta}. Multiple Higgs boson production
processes
are therefore instrumental in this endeavour, not only because they directly depend 
on the Higgs self--couplings, but also because they are
sensitive to possible new heavier states and/or to higher--dimensional
operators \cite{Asakawa:2010xj,Dawson:2012mk,Contino:2012xk,Kribs:2012kz,
Dolan:2012ac,Dolan:2012rv,Gouzevitch:2013qca,Barger:2003rs}. Extracting 
the Higgs self--couplings from collider data is known to be
an arduous task~\cite{Baur:2003gpa,Baur:2003gp}. While the triple Higgs rates 
lie beyond the reach of the LHC capabilities \cite{Plehn:2005nk,Binoth:2006ym}, the prospects
of measuring Higgs pair production are better, but still challenging. 
There are multiple production channels which lead to Higgs boson pairs
at hadron colliders, including vector boson fusion; associated production
with
gauge bosons and heavy quarks; and gluon fusion. The latter is dominant in the SM
and gives an approximate next--to--leading order (NLO) cross section of approximately 35 fb at the 14 TeV LHC \cite{Frederix:2014hta}.
Recent studies \cite{Papaefstathiou:2012qe,Baglio:2012np,Dolan:2012rv,Barr:2013tda,Gouzevitch:2013qca,Li:2013flc,Goertz:2013kp,deLima:2014dta} have
investigated the potential of digging out the di--Higgs signal over its backgrounds in various Higgs decay channels,
among them $\gamma\gamma b\bar{b}$, $\PW^+\PW^-b\bar{b}$, $b\bar{b}\tau^+\tau^-$ and $b\bar{b}b\bar{b}$,
 assisted e.g. by jet substructure techniques.
The current picture is that,   
aside from potential new physics effects, the extraction of the coupling $g_{hhh}$  
will require copious integrated luminosity. Optimistic estimates point towards values 
of $3000$ fb$^{-1}$ at 14 TeV in order to reach an accuracy of $\sim 40\%$ \cite{Barger:2013jfa}. 
The reason is not only the rather modest total rate, but also
the limited sensitivity to
the trilinear coupling $g_{hhh}$ and the large theoretical uncertainties.
With this research program ahead, 
accurate predictions for the total and differential di--Higgs rates are in order --
for the SM and beyond, while more sophisticated experimental analyses are required to explore the precise reach of the LHC.

\medskip{}
In this work we investigate the phenomenological possibilities 
of Higgs pair production via gluon fusion beyond the SM. We resort to the 
2HDM as an illustrative new physics framework, and present results for all seven 
combinations of 2HDM Higgs pair final states. We compute the total and differential
Higgs pair rates at NLO accuracy
in QCD, matched to the {\sc Pythia8}~\cite{Sjostrand:2007gs}
parton shower (PS) with the {\sc MC@NLO} method.  A dedicated reweighting strategy is employed
to improve the NLO calculation beyond 
the infinite top--mass limit, by also employing the exact real--emission matrix elements~\cite{Maltoni:2014eza}.
On the practical side, we carry out our computation
using the MadGraph5$\_${ \sc aMC@NLO } framework \cite{Alwall:2014hca}. 
The results presented here improve upon, and further extend,
the earlier studies in the literature.
To the best of our knowledge, our work provides 
for the first time NLO+PS event samples for these processes, which can be readily used for realistic simulations, 
including those at the detector level once the Higgs bosons are allowed to decay. Accurate predictions
for the differential rates are also important,
as they can be used to identify the distinctive properties of the signal
kinematics and compare them to the backgrounds. 
This is e.g. instrumental in the context of 
jet substructure techniques ~\cite{Dolan:2012rv,Dolan:2012ac}, which  
are known to improve upon the more
traditional search strategies based on inclusive observables.

\medskip{}
The remainder of this paper is organized as follows. In
Section~\ref{sec:model} we begin by outlining the 2HDM
setup, field content, coupling structure, and parameter space constraints. 
Section~\ref{sec:benchmarks} defines a series
of 2HDM benchmark scenarios, which are constructed to cover all representative
phenomenological features of the model. In
Section~\ref{sec:channels} we move on to describe 
the theoretical structure of  
Higgs pair production at leading and next--to--leading order.
Before closing this first part, the technical setup of the calculation is introduced 
in Section \ref{sec:setup}. 
In the second part of the paper we present a comprehensive numerical
analysis of all Higgs pair production channels within the 2HDM.
Total rates are documented and discussed in Section~\ref{sec:rates},
while in Section~\ref{sec:distributions} we focus on the light di--Higgs differential distributions.
Finally, we summarize and conclude in Section \ref{sec:conclusions}.

\section{Phenomenological framework}
\label{sec:pheno}

\begin{table}[t!]
\begin{tabular}{l|l|l}
 & \multicolumn{1}{c|}{Type I} &  \multicolumn{1}{c}{Type II} \\ \hline
 $1+\Delta^{\hzero}_t$ & $\cfrac{\cos\alpha}{\sin\beta} = 1+\xi/\tan\beta -\xi^2/2 + \mathcal{O}(\xi^3)$ & $\cfrac{\cos\alpha}{\sin\beta} = 1+\xi/\tan\beta -\xi^2/2 + \mathcal{O}(\xi^3)$ \\
 $1+\Delta^{\hzero}_b$ & $\cfrac{\cos\alpha}{\sin\beta} = 1+\xi/\tan\beta -\xi^2/2 + \mathcal{O}(\xi^3)$ & $-\cfrac{\sin\alpha}{\cos\beta} = 1-\xi\,\tan\beta -\xi^2/2 + \mathcal{O}(\xi^3)$ \\ \hline
 $1+\Delta^{\Hzero}_t$ & $\cfrac{\sin\alpha}{\sin\beta} = -1/\tan\beta +\xi +\xi^2/(2\tan\beta) + \mathcal{O}(\xi^3) $ & $\cfrac{\sin\alpha}{\sin\beta} = -1/\tan\beta +\xi +\xi^2/(2\tan\beta) + \mathcal{O}(\xi^3) $ \\
 $1+\Delta^{\Hzero}_b$ & $\cfrac{\sin\alpha}{\sin\beta} = -1/\tan\beta +\xi +\xi^2/(2\tan\beta) + \mathcal{O}(\xi^3) $ &$\cfrac{\cos\alpha}{\cos\beta} = \tan\beta +\xi -\xi^2/2\,\tan\beta + \mathcal{O}(\xi^3) $ \\
\end{tabular}
\caption{Heavy--quark Yukawa couplings to the light (heavy)
 \CP--even Higgs 
bosons for type I and type II 2HDM. These are parametrized as a shift
from the SM, according to Eq.~2.2. Their decoupling behavior
is given in terms of the expansion
parameter $\xi \equiv \cos(\beta-\alpha)$ up to $\mathcal{O}(\xi^3)$.\label{tab:yukawas}
}
\end{table}

\subsection{The Two--Higgs--Doublet Model}
\label{sec:model}

The Two--Higgs--Doublet Model (2HDM)~\cite{Branco:2011iw} extends the 
minimal scalar sector of the SM by introducing 
a second $SU(2)_L$ doublet $\Phi_2$ with weak hypercharge
$Y=+1$. This gives rise to an enlarged particle content with five physical
Higgs bosons: one light (resp. heavy) neutral, \CP--even state $\hzero$ (resp. $\Hzero$); 
one neutral, \CP-odd state $\Azero$; and two charged Higgs bosons $\PHiggs^{\pm}$.
Throughout the paper we identify the state $\hzero$ with the Higgs particle
observed at the LHC and fix its mass to $m_{\hzero} = 126$ GeV.
The mixing angle $\alpha$ is introduced to diagonalize the \CP--even squared
mass matrix. Assuming natural flavor conservation~\cite{Glashow:1976nt},
the absence of 
tree--level flavor--changing neutral--current (FCNC) interactions
is protected by a global,
flavor--blind, $Z_2$ discrete symmetry $\Phi_{i} \to (-1)^i \Phi_{i}$.
The latter is approximate up to the soft--breaking mass term
$ \lag_{\text{soft}} \supset m_{12}^2\,\Phi_1^\dagger\,\Phi_2 + \text{h.c.} $
We neglect extra sources of \CP--violation
by considering real mass terms and self--couplings in
the Higgs potential. After electroweak symmetry breaking,
the neutral components of the Higgs doublets
acquire real VEVs, $\braket{\Phi^0_i} = v_i/\sqrt{2}$,
where $v^2 \equiv v_1^2+v_2^2 = G_F^{-1}/\sqrt{2}$. The ratio of the two VEVs
is given by $\tan\beta \equiv v_2/v_1$.
Overall, we are left with 7 free input parameters, which we can sort out as:
\begin{alignat}{9}
 \tan\beta\;, \sin\alpha\;, m_{\hzero}\;, m_{\Hzero}\;, m_{\Azero}\;, m_{\PHiggs^{\pm}}\;, m^2_{12}
 \label{eq:freeinputs}.
\end{alignat}

The convention
$0 \leq \beta-\alpha < \pi$ (with $0 < \beta < \pi/2$) guarantees
that 
the Higgs coupling to the weak gauge bosons $g^{\text{2HDM}}_{hVV} = \sin(\beta-\alpha)\,g^{\text{SM}}_{hVV}$ 
has the same sign in the 2HDM and in the SM. This criterion fixes 
the possible sign ambiguities in the generic parametrization of the model~\cite{Haber:2006ue,Davidson:2005cw}.
\medskip{} 

The
different possible choices of fermion field transformations
under $Z_2$ lead to different Yukawa coupling patterns. We will hereafter
concentrate on the two canonical setups: i) \emph{type--I}, in which 
all fermions couple to just one of the Higgs
  doublets; and ii) \emph{type--II}, where up--type (down--type) fermions couple
  exclusively to $\Phi_2$ ($\Phi_1$). 
The resulting Yukawa couplings deviate from the SM in a way that we
parametrize in the notation of~\cite{Klute:2012pu},
\begin{alignat}{9}
g_{hxx} &\equiv g^h_x  = 
\left( 1 + \Delta^h_x \right) \; g_x^\text{SM} \; .
\label{eq:delta}
\end{alignat}  
  
Analytic expressions for these coupling shifts are provided 
in Table~\ref{tab:yukawas} for both type--I and type--II 2HDMs. 
For further reference, also the trilinear Higgs self--couplings involving
the \CP--even neutral states are quoted below in Table \ref{tab:triple}.

\bigskip{}
Multiple conditions  
place constraints on the parameter space of the model.
On the one hand, unitarity~\cite{\unitarity,Gunion:2002zf}, perturbativity~\cite{Chen:2013kt}
and vacuum stability~\cite{\vacuum} guarantee the 
correct high--energy behavior of the theory. One important
consequence is that the Higgs self--interactions
cannot be arbitrarily large. 
On the other hand, agreement with electroweak precision tests
compresses the allowed mass splitting between the 
heavy scalar fields \cite{\ewpo}, and therefore 
prevents an exceedingly large deviation from the 
(approximate) custodial
$SU(2)$ invariance ~\cite{Veltman:1976rt}. 
Fixing the Higgs mass to $m_{\PHiggs} = 126$~GeV, a global fit to electroweak precision
observables in terms of the oblique parameters $S,T,U$ \cite{Peskin:1990zt}
yields $S = 0.03 \pm 0.01$,
$T = 0.05 \pm 0.12$, and $U = 0.03 \pm 0.10$~\cite{ALEPH:2010aa,Baak:2011ze,Beringer:1900zz}.
Aside from these conditions connected to the structure of the model, the allowed parameter space
shrinks even further as we enforce
compatibility with the average LHC Higgs signal strength \cite{ATLAS-CONF-2014-009,CMS-PAS-HIG-13-005}
and the direct collider mass bounds on the heavy neutral \cite{\massboundsneutral} and charged Higgs
bosons \cite{\massboundscharged}. 
Finally, low--energy heavy flavor physics \cite{\flavor}
and the muon
 $(g-2)_\mu$~ data \cite{Dedes:2001nx,Chang:2000ii,Cheung:2003pw}
place additional indirect constraints on the $(m_{\PHiggs^{\pm}})-\tan\beta$ plane. All these constraints
have been carefully included in our analysis, as we describe in detail below in Section~\ref{sec:benchmarks}.

\begin{table}[tb!]
\begin{tabular}{lll} \hline
$\lambda_{\hzero\hzero\hzero}:$ & 
& $-\cfrac{3}{\sdb}\left[\cfrac{4\cab\cos^2(\beta-\alpha)\,\motd}{\sdb}
- \mhd\,(2\cab\,+ \sda\,\sba)\right] $ \\
& & 
$= 3\mhd + \xi^2\left[\cfrac{9\mhd}{2} - \cfrac{12\motd}{\sdb} \right] + \mathcal{O}(\xi^3)$ \\
$\lambda_{\hzero\hzero\Hzero}:$ &  & $\cfrac{\cba}{\sdb}\,\left[\sda\,(2\mhd + \mHHd) - \cfrac{2\motd}{\sdb}\,(3\sda - \sdb) \right] $ \\
& & $=-\xi\,\left(2\mhd + \mHHd -\cfrac{8\motd}{\sdb} \right) - \cfrac{2\xi^2\,}{\tan 2\beta}\,\left(2\mhd+\mHHd - \cfrac{6\motd}{\sdb} \right) + \mathcal{O}(\xi^3) $ \\
$\lambda_{\Hzero\Hzero\hzero}:$ & & $-\cfrac{\sba}{\sdb}\,\left[\sda\,(\mhd + 2\mHHd)-\cfrac{2\motd}{\sdb}\,(3\,\sda + \sdb) \right] $ \\
 & & $= \mhd + 2\mHHd - \cfrac{4\,\motd}{\sdb} + \cfrac{2\xi}{\tan 2\beta}\,\left(\mhd + 2\mHHd - \cfrac{6\motd}{\sdb}\right) $ \\
 & & $\qquad +\,\xi^2\,\left(\cfrac{14\motd}{\sdb}-\cfrac{5}{2}\,(\mhd+2\mHHd)\right) + \mathcal{O}(\xi^3)$ \\
$\lambda_{\Hzero\Hzero\Hzero}:$ &  & $-\cfrac{3}{\sdb}
\,\left[\mHHd\,(\cba\,\sda - 2\sab) + \cfrac{4\,\motd\,\sab\,\sbad}{\sdb} \right] $ \\
 & & $= \cfrac{-6}{\tan 2\beta}\,\left[\mHHd - \cfrac{2\motd}{\sdb} \right] + \xi\,\left(9\mHHd-\cfrac{12\,\motd}{\sdb} \right)$ \\
 &  & $\qquad + \cfrac{3\xi^2}{\tan 2\beta}\,\left( 3\mHHd - \cfrac{6\motd}{\sdb}\right)  + \mathcal{O}(\xi^3)$ \\ \hline
\end{tabular}
\caption{Triple Higgs self--interactions involving the neutral \CP--even Higgs fields in the 2HDM. 
These are normalized as $\lambda_{hhh} \equiv  i\,v\,g_{hhh}$. 
The Higgs self--coupling in the SM is given by $g_{HHH}^{\text{SM}} = -3im^2_{h}/v$.
Their decoupling behavior
is given in terms of the expansion
parameter $\xi \equiv \cos(\beta-\alpha)$ up to $\mathcal{O}(\xi^3)$.\label{tab:triple}}
\end{table}

\subsection{Benchmarks}
\label{sec:benchmarks}

Phenomenologically viable 2HDM scenarios satisfying all model constraints
and compatible with the LHC data have been extensively scrutinized in the literature \cite{\fits}. 
These studies highlight a preference for
a low--mass, SM--like Higgs field $\hzero$ along with heavier Higgs companions. 
In the decoupling limit \cite{Gunion:2002zf}, the 2HDM can be mapped onto an effective theory,
whose expansion parameter $\xi \equiv \cos(\beta-\alpha) \sim v^2/M^2_{\text{heavy}}$ determines
the hierarchy between the light $m_{\hzero} = \mathcal{O}(v)$ and
the heavy scalar masses $m_{\Hzero} \simeq m_{\Azero} \simeq m_{\PHiggs^{\pm}} \simeq
\mathcal{O}(M_\text{heavy})$ \cite{Gunion:2005ja,Randall:2007as}. 
The decoupling condition $\xi = \cos(\beta-\alpha) \ll 1$ correlates 
the two mixing angles through $\cos\beta \sim \sin\alpha$, which in the 
 limit of large $\tan\beta$ can be expressed by
\begin{alignat}{5}
\sin^2 \alpha \sim \frac{1}{1+ \tan^2 \beta} 
\qquad \text{or} \qquad
\xi \sim \dfrac{2 \tan \beta}{1 + \tan^2 \beta} \; .
\label{eq:xi_tgb}
\end{alignat}
The behavior of the relevant Higgs interactions in the decoupling limit is
explicitly shown
in Tables~\ref{tab:yukawas}-\ref{tab:triple}. 
Notice that even for $\xi \ll 1$ some of the 
couplings may be substantially shifted. 
This behavior appears for instance 
in the $\tan\beta \gg 1$ ($\tan\beta \ll 1$) regimes
as a reflect of the so--called \emph{delayed decoupling} \cite{Haber:2000kq}. 
As for the Yukawa couplings, these shifts may be 
more (in type--I) or less (in type--II) correlated 
within each fermion generation and can lead to enhanced, suppressed, or
even sign--flipped couplings ~\cite{Ferreira:2014naa}. 
In turn, the triple Higgs self--coupling $g_{\hzero\hzero\hzero}$
may be enhanced by up to 100\% above the SM in type--I
models -- while for type--II, the LHC data
favour $g_{\hzero\hzero\hzero} \lesssim g^{\text{SM}}_{HHH}$ with allowed
suppressions up to
$\mathcal{O}(50)\%$ ~\cite{Dumont:2014wha}.
Let us also note the particular decoupling limits  
$g_{\hzero\hzero\hzero} \to g^{\text{SM}}_{HHH}$ and $g_{\Hzero\hzero\hzero} \to 0$  for $\xi \ll 1$. 
The potentially large Higgs self--coupling deviations 
constitute a genuine trait of the 2HDM, with no counterpart in e.g.
the Higgs sector of the MSSM. In the latter case, Supersymmetry relates 
all Higgs self--couplings to the gauge couplings, implying
that their size
becomes restricted.  
The fact that
the conventional decoupling limit $\xi \ll 1$ is not the unique
2HDM setup consistent with a SM--like $\sim 126$ GeV resonance is another
remarkable feature of the model.
In the so--called \emph{alignment limit} \cite{Gunion:2002zf,Craig:2013hca,Carena:2013ooa,Delgado:2013zfa}, 
a SM--like Higgs state can still be made compatible with 
additional Higgs bosons as light as $\sim 200$ GeV \cite{\fits}. Interestingly,
this low--mass region $m_{\Hzero} \lesssim 250$ GeV is also elusive to direct searches -- mainly because of the 
problematic background isolation \cite{Kanemura:2014dea}.

\begin{table}[tb!]
\begin{center}
 \begin{tabular}{|l||rr|rrrr|} \hline \hline
 & $\tan\beta$ & $\alpha/\pi$ & $m_{\Hzero} $ &  $m_{\Azero} $  & $m_{\PHiggs^{\pm}} $   & $m^2_{12} $  \\ \hline \hline
B1 & 1.75 & -0.1872 & 300 & 441 & 442 & 38300  \\
B2 & 1.50 & -0.2162 & 700 & 701 & 670 & 180000 \\
B3 & 2.22 & -0.1397 & 200 & 350 & 350 & 12000 \\
B4 & 1.20 & -0.1760 & 200 & 500 & 500 & -60000 \\
B5 & 20.00 & 0.0000 & 200 & 500 & 500 & 2000 \\
B6 & 10.00 & -0.0382 & 500 & 500 & 500 & 24746 \\
B7 & 10.00 & 0.0323 &  500 & 500 & 500 & 24746 \\ \hline \hline
 \end{tabular}
 \end{center}
 \caption{Parameter choices for the different 2HDM benchmarks used in our study. All
 masses are given in GeV. The lightest Higgs mass is fixed in all cases to $m_{\hzero} = 126$ GeV.}
 \label{tab:benchmarks}
\end{table}

\begin{table} [tb!]
\begin{center}
    \begin{tabular}{ | l || r | r | r | r | r | r |}
    \hline \hline
      & $\hat{g}_{\hzero tt}$ & $\hat{g}_{\hzero bb}$ & $\hat{g}_{\Hzero tt}$ & $\hat{g}_{\Hzero bb}$ & $\hat{g}_{\hzero\hzero\hzero}$  & $\hat{g}_{\Hzero\hzero\hzero}$    \\ \hline \hline
    B1 & 0.958 & 1.118 & -0.639 & 1.677 & 0.956  & -0.317 \\ 
    B2 & 0.935 & 1.132 & -0.755 & 1.403 & 0.592 & -2.058 \\ 
    B3  & 0.993 & 1.035 & -0.466 & 2.204 & 0.999 & -0.019  \\ 
    B4 & 1.108 & 1.108 & -0.684 & -0.684 & 1.324 & -1.542  \\ 
    B5  & 1.001 & 1.001 & 0 & 0 & 0.995 & 0.042  \\ 
    B6 & 0.998 & 1.203 & -0.120 & 9.978 & 0.986 & -0.346   \\ 
    B7  & 0.999 & -1.018 & 0.102 & 9.998 & 0.991 & -0.951  \\  \hline \hline  
   \end{tabular}
\end{center}
\caption{Normalized heavy--quark Yukawa and trilinear Higgs self--couplings for the
different 2HDM benchmarks defined in Table~3. All couplings are normalized to 
their SM counterparts.}
\label{tab:couplings}
\end{table}

\medskip{} 
These rich phenomenological possibilities are captured by the set of 2HDM benchmark scenarios
which we introduce 
in Table~\ref{tab:benchmarks}. We employ them
further down in Section \ref{sec:results} to examine the distinctive
2HDM signatures on the Higgs pair production observables.
They have been constructed in 
agreement with all up--to--date parameter
space constraints, which we have included through an in--house interface of 
the public tools { \sc 2HDMC}~\cite{Eriksson:2009ws},
{ \sc HiggsBounds}~\cite{\higgsbounds}, 
{\sc SuperIso}~\cite{\superiso} and {\sc{HiggsSignals}} \cite{Bechtle:2013xfa,Stal:2013hwa} along with 
additional routines of our own. For the most recent direct heavy Higgs searches \cite{CMS-PAS-HIG-13-025,Aad:2014yja,CMS:2014ipa} not yet
available from {\sc HiggsBounds}, it has been checked explicitly that the benchmarks evade the exclusion bounds. 
In order to better illustrate certain model features,
in some scenarios we tolerate deviations  
slightly beyond $1\sigma$ in the averaged Higgs signal strength.

\medskip{}
In Table~\ref{tab:couplings} we quote the numerical values
for sample Yukawa
and Higgs self--couplings (a selection which is relevant to the light Higgs pair production) 
for all seven 2HDM benchmarks 
defined in Table~\ref{tab:benchmarks}. 
All couplings
are normalized to their SM counterparts, as denoted
by $\hat{g}_{hxx} \equiv g^{\text{2HDM}}_{hxx}/g^{\text{SM}}_{Hxx}$,
where $H$ stands for the SM Higgs boson. The heavy Higgs trilinear coupling
is normalized as $\hat{g}_{\Hzero \hzero\hzero} \equiv g_{\Hzero\hzero\hzero}/g^{\text{SM}}_{HHH}$.

\medskip{}
The key physics properties of the different 2HDM scenarios 
can be summarized as follows:

\begin{itemize}
 \item{\textbf{B1: Moderate mass hierarchy - }
   taken from benchmark H1 in Ref.~\cite{Baglio:2014nea}. It 
   corresponds to a type II 2HDM with moderate (viz. $300-500$ GeV) heavy Higgs masses. 
  The small values of $\tan\beta$ and $\xi \equiv \cos(\beta-\alpha)$
  guarantee that all light Higgs boson couplings remain very close to 
  the SM -- with an $\mathcal{O}(5)\%$ suppression in the triple Higgs self coupling
  and an $\mathcal{O}(10)\%$ enhancement in the bottom Yukawa.    
  This is also the reason why this benchmark evades the recent 
  CMS \cite{CMS-PAS-HIG-13-025} and ATLAS limits \cite{Aad:2014yja}. 
  As we will show in Section~\ref{sec:results}, 
  the major new physics effects
  in this case originate from the resonant heavy Higgs--mediated contribution $gg \to \Hzero \to \hzero\hzero$. These effects
  are particularly enhanced here due to the dominant heavy Higgs cascade decay $\Hzero \to \hzero\hzero$,
  whose branching fraction is $\text{BR}_{\hzero\hzero} \simeq 0.6$. 
  }
 \item{\textbf{B2: Large mass hierarchy - }
  taken from benchmark a.1 again in Ref.~\cite{Baglio:2014nea}.
  At variance with the B1 scenario above, all of the additional Higgs bosons
  have in this case masses of $\mathcal{O}(700)$ GeV, so that they effectively decouple. The very 
  large $Z_2$ soft--breaking term $m^2_{12}$ is responsible for the $\mathcal{O}(40)\%$ suppression
  of the trilinear Higgs self--coupling $g_{\hzero\hzero\hzero}$, while at the same
  time it enlarges $g_{\Hzero\hzero\hzero}$. 
 }
 \item{\textbf{B3: Non--resonant SM limit -}
 in which the relatively light \CP--even Higgs companion  
 $m_{\Hzero} = 200$ GeV $< 2\,m_{\hzero}$ precludes
 the resonant production $gg \to \Hzero \to \hzero\hzero$. The parameter choice is inspired
 by the benchmarks of class [c] from Ref.~\cite{Baglio:2014nea}. They are characterized
 by a rather light Higgs spectrum, along with moderate values for $\tan\beta$ and $m^2_{12}$.
 The Yukawa interactions are once again fixed according to
 a type--II setup. The resulting coupling patterns in this case
 are SM--like for the lightest Higgs field, with mild
 deviations not larger than $1\%$. For the heavy neutral field
 $\Hzero$ we find a suppressed (resp. enhanced) 
 Yukawa coupling to the top (resp. bottom) with opposite signs,
 and a strongly reduced trilinear coupling $g_{\Hzero\hzero\hzero}$. 
 }
 \item{\textbf{B4: Enhanced triple Higgs self--coupling --}
  we assume 
  i) type--I Yukawa couplings, for which the LHC Higgs signal strength
  imposes weaker constraints; 
  ii) a small $\tan\beta$ value; iii) a value of $\xi$ for which
  the Higgs coupling to the gauge bosons $g_{hVV} \sim \sin(\beta-\alpha) \sim 1 - \xi^2/2$
  is reduced by $\sim \mathcal{O}(1)\%$; 
  iv) a soft--breaking mass term close to the unitarity limit.
  Such parameter setup leads to
  an $\mathcal{O}(30)\%$ enhancement for the light Higgs triple self--coupling, and of 
  $\mathcal{O}(10)\%$ for the heavy quark Yukawas. The negative
  sign  $m_{12}^2 < 0$ protects the stability of the vacuum. The relatively low mass $m_{\Hzero} < 2 m_{\hzero}$
  prevents the resonant heavy Higgs production and allows to  
  better access the genuine model--specific imprints on the di--Higgs production observables.}

 \item{\textbf{B5: Fermiophobic heavy Higgs --}
 a situation which can only be accomplished in type--I models
 for $\sin\alpha =0$. On the one hand, 
 compatibility with the LHC Higgs signal strength enforces $\xi \ll 1$, meaning
 that these scenarios are only viable at large $\tan\beta$, according to Eq.~\eqref{eq:xi_tgb}. The value of
 $m_{12}^2$ is tailored to fulfill the vacuum stability 
 and unitarity bounds, which become very tight at large $\tan\beta$.
 Notice that for $\xi \ll 1$ 
 the (relatively low--mass) heavy Higgs $\Hzero$ hardly
couples to the gauge bosons and the light Higgs $\hzero$, 
while by construction it cannot couple 
to any fermion either. 
Consequently, in this case there is no heavy Higgs contribution to the light di--Higgs production 
$gg \to \Hzero \to \hzero\hzero$.} 

 \item{\textbf{B6 and B7: Enhanced and sign--flipped bottom Yukawa -- }
 both instances are possible in type--II models at large $\tan\beta$,
where the bottom Yukawa coupling $g_{\hzero bb} = (1-\xi\tan\beta)\,g_{Hbb}^{\text{SM}}$ may yield either
 $g_{\hzero bb} \simeq - g_{Hbb}^{\text{SM}}$
 or $g_{\hzero bb} > g_{Hbb}^{\text{SM}}$, in correspondence to
the two branches $\sin\alpha > 0$ and $\sin\alpha < 0$
 along the decoupling condition of Eq.~\eqref{eq:xi_tgb}.
 The possibility of a strongly modified 
 bottom Yukawa, with all of the remaining
 couplings being SM--like, is a trademark property of the 2HDM \cite{Ferreira:2014naa}, and relies
on the \emph{delayed decoupling} behavior mentioned earlier~\cite{Haber:2000kq}. 
 A similar mechanism for $\tan\beta < 1$ should in principle lead to a sign--reverted top Yukawa,
 although this situation is in practice disfavored by the flavor constraints.}
\end{itemize}

\subsection{Higgs pair production in the 2HDM}
\label{sec:channels}

The pair production of Higgs bosons at hadron colliders
can proceed via weak gauge boson fusion ~\cite{Dobrovolskaya:1990kx,Dicus:1987ez,Abbasabadi:1988bk,Abbasabadi:1988ja,Eboli:1987dy,Keung:1987nw}, 
double Higgs--strahlung off the W and Z
bosons~\cite{Barger:1988jk}, and 
gluon fusion~\cite{Glover:1988rg,Plehn:1996wb,Eboli:1987dy,Dicus:1987ic}. 
Because of the large gluon luminosity in the high--energy proton beams, 
the gluon gluon fusion channel dominates. Predictions
for the SM are known at
NLO \cite{Dawson:1998py} and NNLO \cite{deFlorian:2013uza} in QCD, 
both in the infinite top--mass effective theory. Further studies have reported 
on the subleading $\mathcal{O}(1/m^2_t)$
terms \cite{Grigo:2013rya}, threshold resummation \cite{Shao:2013bz}, 
as well as on gluon fusion results
merged to one jet \cite{Maierhofer:2013sha}. More recently,  
predictions for all Higgs pair production channels at NLO and matched to parton showers 
have been presented in Ref.~\cite{Frederix:2014hta}. 
These studies conclude that higher--order effects are large, especially for the dominant gluon fusion channel, 
and that including them significantly reduces
the theoretical uncertainties. 

\smallskip{}
In the context of new physics, 
double Higgs production has been addressed in a variety of models, 
for instance the MSSM \cite{Plehn:1996wb,Djouadi:1999rca}, the NMSSM~\cite{Cao:2013si,Ellwanger:2013ova,Nhung:2013lpa}, the 2HDM~\cite{Arhrib:2008pw,Arhrib:2009hc,Asakawa:2010xj}, 
extended colored sectors~\cite{Dawson:2012mk,Kribs:2012kz,Chen:2014xwa},
Little Higgs~\cite{Dib:2005re,Wang:2007zx}, Higgs portal~\cite{Dolan:2012ac,No:2013wsa}, a flavor symmetry model \cite{Berger:2014gga}
and Composite Higgs models~\cite{Grober:2010yv,Gillioz:2012se}. Dedicated studies on the 
charged Higgs pair case have been presented e.g. in Refs.~\cite{Krause:1997rc,Brein:1999sy}.
Model--independent approaches
based on anomalous couplings and effective operators
have been considered as well~\cite{Pierce:2006dh,Contino:2012xk,Liu:2013woa}. 

\begin{figure}[tb!]
\begin{center}
\includegraphics[width=0.7\textwidth]{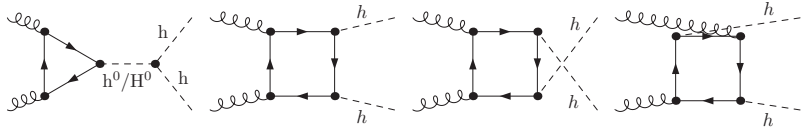}
\end{center}
\caption{Generic Feynman diagrams describing the 
production of neutral Higgs boson pairs ($h = \hzero, \Hzero, \Azero$) in the 2HDM through gluon fusion
at leading order. The Feynman diagrams
have been generated using {\sc FeynArts.sty} \cite{Hahn:2000kx}.}
\label{fig:cpeven}
\end{figure}

\begin{figure}[tb!]
\begin{center}
\includegraphics[width=0.7\textwidth]{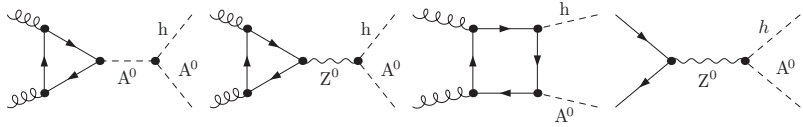}
\end{center}
\caption{Generic Feynman diagrams describing the production of mixed  
\CP--even / 
\CP--odd neutral Higgs boson pairs [$\hzero\Azero, \Hzero\Azero$] in the 2HDM at leading order.
We separately show the two possible partonic initial--states
i) gluon fusion (left, center); ii) $q\bar{q}$ annihilation (right--most).}
\label{fig:hA}
\end{figure}

\bigskip{}
In this study we focus on the Higgs pair production in the 2HDM, and consider the seven possible
final--state double Higgs combinations. These may be sorted out into the following three categories:
\begin{enumerate}
 \item {Neutral Higgs boson pairs: $\hzero\hzero,\hzero\Hzero,\Hzero\Hzero,\Azero\Azero$.}
 \item{Mixed $\CP$--even/$\CP$--odd neutral Higgs boson pairs: $\hzero\Azero,\Hzero\Azero$. }
 \item{Charged Higgs boson pairs: $\PHiggs^+\PHiggs^-$.}
\end{enumerate}

There are two leading--order (LO) mechanisms 
contributing to the gluon fusion di--Higgs channels $pp(gg) \to hh$ (with $h = \hzero, \Hzero, \Azero$), whose generic
Feynman diagrams we display in Figure~\ref{fig:cpeven}. These
correspond to: 
\begin{enumerate}
\item{Triangle topologies, which give rise to $\mathcal{O}(G_F\,\alpha_s\,\hat{g}_q)$
contributions 
through the s--channel exchange of (at least) one neutral Higgs boson. The Higgs boson couples to the gluons
via the usual heavy--quark loops.}
\item{Box topologies, which contribute at 
$\mathcal{O}(G_F\,\alpha_s\,\hat{g}_q^2)$ through the virtual heavy quark exchange.}
\end{enumerate}

The normalized heavy quark Yukawas  $\hat{g}_q \equiv g_q/g_q^{\text{SM}}$ are
related to the 2HDM coupling shifts through Eq.~\eqref{eq:delta}
and Table~\ref{tab:yukawas}. 
While the triangle topologies 
have a linear dependence in $\hat{g}_q$
and in the Higgs self--coupling, the boxes are quadratic in the former. 
In the SM, the two topologies interfere destructively. This effect is particularly strong
near the threshold \cite{Li:2013rra} and explains in part 
why the total rates are quite modest. One additional 
destructive interference arises between the top and bottom--mediated
loops, although the bottom quark effects are very small in the SM.

For category 2), we have additional tree--level 
contributions
for which a quark-antiquark pair annihilates into a virtual $\PZ$--boson (cf. the right--most
diagram in Figure~\ref{fig:hA}).
For the charged Higgs boson pairs of category 3), also the photon exchange
from the $q\bar{q}$--annihilation contributes. 
The relative size of these $q\bar{q}$--initiated subchannels is quantified in Section~\ref{sec:rates}.

\medskip{}
To gain further insight into the structure of the (loop--induced)
gluon fusion mechanism, let us focus on the neutral \CP--even Higgs
pairs of category 1). 
The partonic cross--section at leading order may be written as

\begin{alignat}{9}
 \cfrac{d{\hat{\sigma}(gg \to h_ih_j)}}{d\hat{t}} = 
 c_{ij}\,\cfrac{G_F^2\,\alpha^2_s}{2^8\,(2\pi)^3}\, & \sum_{\substack{q=t,b \\ h = \hzero,\Hzero}}\Bigg{\{} 
 \left| \left(\cfrac{g_{hqq}\,v}{m_q}\,\times\,\cfrac{g_{h\,h_i\,h_j} \, v}{s-m^2_{h} + i\,m_h\,\Gamma_{h}}\right)\,F_{\bigtriangleup} 
  +\cfrac{g_{h_i\,qq}\,g_{h_j\,qq}\,v^2}{m_q^2}\,F_{\Box}
 \right|^2 
  \notag \\
 & \qquad \qquad + \left|\cfrac{g_{h_i\,qq}\,g_{h_j\,qq}\,v^2}{m_q^2}\,G_{\Box}\right|^2
 \Bigg{\}}
 \label{eq:lo}.
\end{alignat}

\noindent The functions $F_{\bigtriangleup}, F_{\Box}$ and $G_{\Box}$
denote the one--loop gauge--invariant form factors from the triangle and the box
contributions, and depend on the Mandelstam kinematical invariants 
and the quark and Higgs masses. The variable $\hat{t}$ corresponds to the momentum transfer squared from one of the incoming
initial--state gluons
to one of the Higgs bosons in the final state.
The sum over the neutral \CP--even Higgs fields
$h = \hzero,\Hzero$ accounts for the respective Higgs--mediated triangle topologies. 
Notice that the s--channel exchange of the \CP--odd Higgs $\Azero$ is forbidden by \CP--conservation.
The model--dependent Yukawa couplings are related to their SM counterparts 
through the coupling shifts quoted in Table~\ref{tab:yukawas}. 
As for the Higgs self--interactions (cf. Table~\ref{tab:triple}),
we normalize them as $\lambda_{hhh} \equiv  i\,v\,g_{hhh}$,
with the SM Higgs self--coupling being $g_{HHH}^{\text{SM}} = -3im^2_{H}/v$. 
Finally, 
the symmetry factor $c_{ij} = 1/2\,(1)$ for $i=j \, (i\neq j)$ properly 
accounts for the cases with identical (different) final--state particles. 
The overall normalization
of Eq.~\eqref{eq:lo} is consistent with the notation of Ref.~\cite{Plehn:1996wb}. 
The corresponding hadronic cross section is obtained by convoluting Eq.~\eqref{eq:lo}
with the gluon luminosities.

\medskip{}
The two box form factors $F_{\Box},G_{\Box}$ correspond to the two 
possible S--wave and D--wave contributions. These 
are linked to equal (resp. opposite) gluon helicities, which respectively
add to a total angular momentum $J_z = 0$ (resp. 
$J_z = 2$) along the collision axis. Instead, the triangle diagrams only contribute
for $J_z = 0$ and hence give rise to a single form factor $F_{\bigtriangleup}$ . 
The large (resp. low) mass limits of the loop form factors read~\cite{Plehn:1996wb}:
\begin{alignat}{2}
 & F_{\bigtriangleup}^{t} = \frac{2}{3} + \mathcal{O}(\hat{s}/m_t^2); \qquad &&    F_{\bigtriangleup}^{b} = -\frac{m_b^2}{\hat{s}}\,\left[\log\left(\cfrac{m_b^2}{\hat{s}}\right)  + i\pi \right]^2 + \mathcal{O}(m_b^2/\hat{s}) \notag \\
 & F_{\Box}^t=-\frac{2}{3} + \mathcal{O}(\hat{s}/m_t^2); \qquad &&   F_{\Box}^b=\mathcal{O}(m_b^2/\hat{s}) \notag \\
 & G_{\Box}^t = \mathcal{O}(\hat{s}/m_t^2); \qquad &&   G_{\Box}^b=\mathcal{O}(m_b^2/\hat{s}) 
  \label{eq:ffactor-limits}
\end{alignat}
The opposite signs for these form factors reflect the 
negative interference patterns between i) boxes and triangles; ii) top and bottom--mediated
loops.

\medskip{}
The Higgs pair total rates and distribution shapes
are determined by the 
size and relative signs
of these different loop--induced contributions, and the way
they interplay  according to Eq.~\eqref{eq:lo}.
The results are thereby
sensitive to the underlying 2HDM dynamics. For instance,
sign--flipped couplings may revert  
the partial cancellation between the triangle and the box
loops. The possible variations in the Yukawa
and the Higgs self--couplings may
either enhance or suppress these interference patterns.
If kinematically allowed, 
the on--shell production of a heavy neutral \CP--even Higgs through
the heavy top triangle,
followed by the cascade decay $\Hzero \to \hzero\hzero$, will overwhelm
the SM expectations. 
All these cases will be examined in Section~\ref{sec:results}
with the help of the different 2HDM benchmark scenarios defined in Table ~\ref{tab:benchmarks}.

\subsection{Next--to--leading order corrections}
\label{sec:nlo}

\begin{figure}[tb!]
\begin{center}
\includegraphics[width=0.7\textwidth]{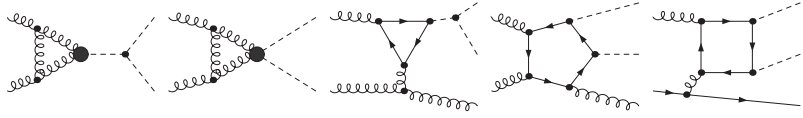}
\end{center}
\caption{Sample Feynman diagrams describing
the production of neutral Higgs pairs ($h = \hzero, \Hzero$) via gluon fusion
at next--to--leading order in the 2HDM. The shaded blobs denote the effective Higgs couplings
to the gluons
in the HEFT approach.}
\label{fig:nlo}
\end{figure}

The NLO QCD corrections arise at $\mathcal{O}(\alpha_s^3)$ and are obviously
linked to the color charges of the initial partons. They originate
from i) virtual gluon exchange; and ii) light parton radiation. The 
NLO virtual corrections to $gg \to hh$ are intrinsically
a two--loop effect, not accessible by current loop calculation techniques. 
As an alternative, Higgs pair production studies at NLO and NNLO~\cite{Dawson:1998py,deFlorian:2013uza} resort to
an Effective Field Theory
framework. In the so--called Higgs Effective Field Theory (HEFT) approach, 
the effective three--point and four--point Higgs boson interactions with the gluons are obtained
by integrating out the top--quark. These effective interactions, 
which are connected to the low--energy theorems \cite{Ellis:1975ap,Shifman:1979eb,Kniehl:1995tn},
can be written in terms of higher--dimensional operators
and cast into the Lagrangian:
\begin{alignat}{9}
 \lag 
  \supset\, \cfrac{1}{3}\,\cfrac{\alpha_s\,(1+\Delta^h_t)}{4\pi v}\,G^{\mu\nu A}\,G^A_{\mu\nu}\,h -
 \cfrac{1}{6}\,\cfrac{\alpha_s\,(1+\Delta^h_t)^2}{4\pi v^2}\,G^{\mu\nu A}\,G^A_{\mu\nu}\,hh
 \label{eq:LET},
\end{alignat}

\noindent where $G^A_{\mu\nu}$ denotes the gluon field--strength tensor while
$h = \hzero,\Hzero$. Similar expressions may be derived for the gluon
couplings to the pseudoscalar Higgs, cf. e.g. \cite{Dawson:1998py}.
The $(1 + \Delta^h_t)$ factor (cf. Table~\ref{tab:yukawas})
accounts for the model--dependent shifts to the effective $ggh$ and $gghh$ interactions.
The relative sign between the two couplings constitutes the low--energy
equivalent of the destructive interference between the triangle and the box topologies.

\medskip{}
A selection of the Feynman
diagrams describing the relevant NLO QCD effects are illustrated in Figure~\ref{fig:nlo}. 
The NLO virtual corrections in the HEFT
picture are obtained from one--loop topologies, as shown
by the first two Feynman diagrams of Figure~\ref{fig:nlo}.
The shaded blobs denote the effective
three--point and four--point Higgs couplings to the gluons. 
In our calculation, these HEFT virtual
corrections are combined with the exact $2\to 3$ real--emission amplitudes
(cf. the right--most diagrams in Figure~\ref{fig:nlo})
as an alternative to the more traditional approach,
which only uses the exact LO result to improve upon the infinite top--mass limit.  
Further details are provided in Section~\ref{sec:setup}. 
Aside from the leading--order $gg$--initiated partonic subchannel,
notice that for the NLO real--emission contributions
also the mixed $qg$ fusion channels open up.

\section{Calculation setup}
\label{sec:setup}

In this section we describe in detail the technical setup of our calculation. 
As already mentioned,
one important aspect is the treatment of the NLO corrections to the gluon fusion channels.
Given that this production mechanism is loop--induced at LO, two
ingredients would be needed for an exact NLO calculation: 
i) the one--loop $2\to 3$ real--emission amplitudes;
ii) the two--loop $2\to 2$ virtual correction amplitudes.  
While the former can be calculated by means of standard techniques, the latter are 
still beyond the reach of the state--of--the--art calculations. 
On the other hand,
the HEFT description from Eq.~\eqref{eq:LET} relies on the 
infinite top--mass approximation and thereby has a limited validity. 
It is in fact well--known that it
provides only a rough estimate of total rates \cite{Dawson:2012mk}, while
it poorly reproduces the kinematical distributions~\cite{Baur:2002rb,Dolan:2012rv}. 
The usual approach in higher--order studies is 
to extract the QCD corrections from the HEFT, and then employ the exact one--loop
LO amplitudes to reweigh the HEFT
virtual- and real--emission matrix elements.
This conventional strategy,
as implemented in {\sc HPAIR}~\cite{Plehn:1996wb,Dawson:1998py},
will be hereafter referred to as the ``born--improved'' approach.
One important point in this procedure is that 
the reweighting
of the $2\to 3$ real--emission part is based
on its factorization into the $2\to 2$ LO matrix element times a global factor.

As an alternative, in this work we include in addition
the ``loop-improved'' approach which was first presented
in \cite{Frederix:2014hta}. 
For completeness, here we describe the method briefly, while a detailed discussion  is provided 
in ~\cite{Maltoni:2014eza}.
In the ``loop-improved'' calculation we include
the exact one--loop results
not only for the $2 \to 2$ LO amplitudes,
but also for the NLO $2 \to 3$ real-emission matrix elements. 
Therefore, the only approximation
we make at the amplitude level concerns the finite part of the NLO virtual corrections which,
in the absence of the exact two--loop calculation, is first taken from the 
one--loop HEFT results, and then reweighted with the exact one--loop LO matrix elements. 
Including the exact one--loop $2 \to 3$ matrix elements 
provides a more accurate 
description of the tails of the distributions, e.g. at large Higgs boson pair transverse momentum 
$p_T(hh)$. In these phase space regions, in which 
hard parton emissions take place, the factorization of 
the $2 \to 3$ real--emission amplitudes into 
the $2 \to 2$ LO amplitudes, as implicit in the ``born--improved'' approach, 
cannot accurately describe the hard parton kinematics.
In Section~\ref{sec:results}, we compare the Higgs pair total rate predictions
obtained within the ``born--improved''
and the ``loop--improved'' methods, while for the differential rates 
we concentrate on the ``loop-improved'' results. 
Despite the upgraded treatment of hard real emission,  
our ``loop--improved'' method still relies on the one--loop 
HEFT virtual corrections in place of the exact two--loop results.
The impact of these exact two--loop results cannot be quantified 
until they become available. 
The validity of the approximation is nonetheless limited, not only because the top quark mass is not 
parametrically large as compared to the other relevant scales of the process,
but also because of the enhanced bottom--quark contributions which
are possible in the 2HDM. 
Given that 
the low--energy theorems do not hold for bottom--mediated loops, 
we expect the HEFT virtual results 
for certain 2HDM configurations to be less reliable than in the SM case.
Let us also note that, for the $gg$--induced production 
of the mixed \CP-- even/\CP-- odd pairs, the NLO virtual corrections
must again be approximated by the one--loop HEFT
results. As for the s--channel $\PZ$--mediated contributions in the latter cases, 
we use the same factorized form as for the $\Azero$--mediated ones,
which is exact in the $m_t\to \infty, m_b\to 0$ limit as discussed in Ref.~\cite{Dawson:1998py}. 
Concerning the NLO real--emission corrections, in all these channels we include the full set of non--zero Feynman 
diagrams contributing to $g g \to g\hzero\Azero/\Hzero\Azero$, among them the box diagrams of the form $g g \to g Z^* \to g\hzero\Azero/\Hzero\Azero$.
The latter were not included in the analysis of Ref.~\cite{Dawson:1998py}, as they cannot be factorized into the LO amplitude and a universal correction factor.
These contributions are infrared and collinear--finite and are expected to be small, but we nevertheless calculate them for completeness.
Similar approximations are made for the charged Higgs pair case.

\medskip{}
From the technical point of view, 
our results are obtained within the \linebreak MadGraph5$\_${ \sc aMC@NLO} framework, which allows the calculation of LO and 
NLO cross sections at a fully differential level. In the MadGraph5$\_${ \sc aMC@NLO} setup, NLO computations are carried out 
using two independent modules:
i) {\sc MadFKS} \cite{Frederix:2009yq}, which takes care of the Born and 
the real--emission amplitudes, subtracts
the infrared singularities according to the FKS prescription \cite{Frixione:1995ms,Frixione:1997np}, and generates the parton shower
subtraction terms required by the MC@NLO method \cite{Frixione:2002ik}; ii) {\sc MadLoop}~\cite{Hirschi:2011pa}, 
which handles the calculation of the one--loop matrix elements
using the {\sc OPP} integrand--reduction method~\cite{Ossola:2006us} (as
implemented in {\sc CutTools}~\cite{Ossola:2007ax}) and the {\sc OpenLoops}
method~\cite{Cascioli:2011va}. 
For the 2HDM Higgs pair processes which are tree--level at LO, 
such as the $q\bar{q}\rightarrow \hzero\Azero$ subchannels, the NLO results may be obtained
automatically within MadGraph5$\_${ \sc aMC@NLO}. At variance, for 
the loop--induced gluon fusion channel events are generated at NLO
using the HEFT results and then reweighted  
using the exact $2 \to 2 $ and $2 \to 3$ one--loop amplitudes, by means of
a separate reweighting routine.
All one--loop matrix elements are computed by {\sc MadLoop}~\cite{Hirschi:2011pa}. 
The final results are in all cases
fully differential, so that they can be used to obtain any distribution at will, 
after matching with the parton shower. 

\medskip{}
As for the 2HDM implementation, 
we use the 2HDM@NLO model obtained with the package {\sc NLOCT} \cite{Degrande:2014vpa}.
The model is based on the {\sc FeynRules} \cite{Alloul:2013bka} and {\sc UFO } \cite{Degrande:2011ua} frameworks. It includes all relevant UV counterterms and R2 vertices needed for the {\sc MadLoop} calculation, and 
allows to compute tree--level and one--loop 
processes within a completely general 2HDM setup. 
The input parameters for the different benchmarks 
are submitted to MadGraph5$\_${ \sc aMC@NLO} \cite{Alwall:2014hca}
through parameter cards in the standard {\sc MadGraph} setup. 
The latter are constructed with the help of an in--house
modification of the public calculator {\sc 2HDMC} \cite{Eriksson:2009ws}.

\medskip{}
The heavy--quark (pole) masses are set to their
current best--average values, $m_t = 173.07$ GeV and $m_b = 4.78$ GeV~\cite{Beringer:1900zz}. 
The lightest neutral \CP--even mass--eigenstate of the 2HDM 
is identified with the SM Higgs boson, 
with a mass $m_{\hzero} = 126$ GeV. The LHC center--of-mass energy is fixed to $\sqrt{S} = 14$ TeV.
All particle widths are set to zero in the loop propagators, while the 
s--channel Higgs boson widths are computed with {\sc 2HDMC} and included in the
MadGraph5$\_${ \sc aMC@NLO} parameter cards. Finite width effects in the heavy quark loop propagators, which can be handled by MadGraph5$\_${ \sc aMC@NLO} via the complex mass scheme, are beyond the scope of this paper
and not taken into account.

\medskip{}
Parton distribution functions (PDFs) are evaluated using the MSTW2008 sets with four active
flavors at LO and NLO consistently~\cite{Martin:2009iq}. Therefore, the $b\bar{b}$--initiated
channels are not included in our calculation.
The MadGraph5$\_${ \sc aMC@NLO } framework allows fully flexible renormalization
and factorization scale choices. For all results that follow in this study, 
we set common factorisation and renormalisation scales,
fixing their central value to 
half the invariant mass of the Higgs pair $\mu_R^0 = \mu_F^0 = m_{h_i\,h_j}/2$, where $ij$ account for all possible 
final--state combinations. This scale choice has been proved to yield
perturbatively stable results for the Higgs boson pair production at NLO \cite{Frederix:2014hta}, and to behave very similarly
with respect
to alternative scale settings used in the literature (cf. e.g. Refs.~\cite{Dawson:1998py,deFlorian:2013jea}).
The strong coupling constant is evaluated at the
renormalization scale $\alpha_s(\mu_R^0)$ along with the PDFs.
The scale and PDF uncertainties are generated automatically and at no extra computational cost
within MadGraph5$\_${ \sc aMC@NLO}, following the reweighting prescription of \cite{Frederix:2011ss}. In our analysis we 
vary independently the scales in the range $ \mu^0/2 <\mu^0_R,\mu^0_F <2\mu^0$.  
PDF uncertainties at the 68\% C.L. are extracted by following the prescription 
given by the MSTW collaboration~\cite{Martin:2009iq}. 
{\sc Pythia8}~\cite{Sjostrand:2007gs} is used for parton shower and hadronisation.
The matching to the {\sc Pythia8} parton shower (virtually ordered, and $p_T$--ordered for
processes with no final--state radiation, such as the production of Higgs pairs) is also
automated within MadGraph5$\_${ \sc aMC@NLO }.
For conciseness, only observables linked to the final--state Higgs pair are shown in
the following. This implies that the Higgs bosons are kept stable at the parton shower stage.
 
\medskip{}
The dedicated codes for the calculation of the gluon fusion channels can be downloaded from~\cite{web}. This website contains in addition a 
selection of standalone codes for different Higgs pair production processes in the
SM and beyond.

\section{Numerical results}
\label{sec:results}

\subsection{Total rates}
\label{sec:rates}

We report
on the 2HDM predictions
for the Higgs pair total cross sections at the LHC. 
Our results are documented in Tables~\ref{tab:B1table}--\ref{tab:b5b6b7}.
Characteristic model--dependent features
are highlighted
by the different benchmarks defined in Section~\ref{sec:benchmarks}.
For the sake of brevity, we select a 
representative subset of them (B1 -- B3)  
to compute the total rates for all seven combinations of 2HDM Higgs pair
final states. The corresponding results are shown in Tables~\ref{tab:B1table}--\ref{tab:B3qqtable}. 
For the remaining parameter choices (B4 -- B7), we concentrate on the light di--Higgs production $\hzero\hzero$ (cf. Table~\ref{tab:b5b6b7}).
In addition to the total rate predictions for the (leading) gluon fusion mechanism,
Tables~\ref{tab:B1qqtable}, \ref{tab:B2qqtable} and ~\ref{tab:B3qqtable} include the contribution from 
the additional
$q\bar{q}$--initiated channels. The latter only feature for the
mixed neutral Higgs pairs ($\hzero\Azero, \Hzero\Azero$) and the charged Higgs pairs $\PHiggs^+\PHiggs^-$, and 
occur already at tree--level through the s--channel $\PZ^0$ boson exchange --
as well as through photon exchange in the charged Higgs case.
We show in all cases the total rate predictions
at LO and NLO accuracy in QCD. For the gluon fusion channels, we explicitly
distinguish between the two reweighting schemes in use, as described earlier in Section~\ref{sec:setup}:
i) the NLO--``loop--improved'', with both the exact $2 \to 2$ LO matrix elements and the exact 
one--loop $2\to 3$ real--emission matrix elements used to reweight the HEFT results; ii)
the NLO--``born--improved'', in which only the exact one--loop $2\to 2$ LO matrix elements are used for
reweighting. The total rate central values 
are shown along with the
scale and PDF uncertainties, whose upper (lower) bands are 
readily evaluated in MadGraph5$\_${ \sc aMC@NLO}. The size of the NLO corrections
is quantified in terms of the $K$--factor, provided in the right--most
column of the tables. The latter
is defined consistently through 
$K \equiv \sigma^{\text{NLO}}/\sigma^{\text{LO}}$, where 
the NLO cross section for the $gg$--initiated channels stands for the ``loop--improved'' result.
For the $q\bar{q}$--initiated channels, the NLO prediction corresponds to the exact result.

\begin{table}[tb!]
\begin{center}
    \begin{tabular}{ | c || r | r | r | r |}
   \multicolumn{5}{l}{\textbf{Benchmark B1}} \\ \hline \hline
     \textbf{$gg$--channels} & LO &NLO--loop improved &NLO--Born improved  & $K$--factor \\ \hline \hline
    $\hzero\hzero$ & 1480$^{+29.8+1.5\%}_{-21.6-2.1\%}$ & 2400$^{+17.9+1.6\%}_{-14.4-1.9 \%}$ & 2500$^{+19.2+1.6\%}_{-15.0-1.9 \%}$ & 1.62 \\ \hline
    $\hzero\Hzero$ & 10.5$^{+33.5+2.3\%}_{-23.5-2.5\%}$ & 16.1$^{+15.2+2.3\%}_{-13.7-2.8 \%}$  & 17.9$^{+18.5+2.3\%}_{-15.3-2.8 \%}$  & 1.54 \\ \hline
    $\Hzero\Hzero$ & 0.550$^{+35.3+2.8\%}_{-24.4-2.9\%}$  & 0.859$^{+14.7+2.8\%}_{-13.8-3.6 \%}$  & 0.936$^{+17.6+2.8\%}_{-15.3-3.5 \%}$  & 1.56   \\ \hline
    $\hzero\Azero$ & 5.22$^{+34.4+2.5\%}_{-23.9-2.7\%}$  & 8.68$^{+17.1+2.5\%}_{-14.8-3.1 \%}$ & 8.90$^{+17.9+2.5\%}_{-15.2-3.1 \%}$  & 1.66 \\ \hline
    $\Hzero\Azero$ & 0.457$^{+36.4+3.1\%}_{-24.9-3.2\%}$& 0.727$^{+15.2+3.3\%}_{-14.3-4.1 \%}$   & 0.798$^{+17.8+3.2\%}_{-15.5-4.0 \%}$ & 1.59  \\ \hline
    $\Azero\Azero$ & 0.221 $^{+37.3+3.4\%}_{-25.4-3.5\%}$& 0.352$^{+14.8+3.7\%}_{-14.2-4.6 \%}$ &  0.382$^{+17.4+3.7\%}_{-15.5-4.6 \%}$ & 1.59 \\ \hline
    $\PHiggs^+\PHiggs^-$ & 0.321$^{+37.3+3.4\%}_{-25.4-3.5 \%}$ & 0.531$^{+16.1+3.7\%}_{-14.9-4.6 \%}$ &  0.559$^{+17.7+3.7\%}_{-15.7-4.6 \%}$ & 1.65 \\  \hline\hline  
   \end{tabular}
\end{center}
\caption{Total Higgs pair cross sections via gluon fusion $pp(gg) \to h_i\,h_j$ (in fb)
for all seven final--state combinations within the 2HDM. The rates
are computed at LO and NLO, including the QCD corrections within either the
``loop--improved'' or
``born--improved'' approach. 
The associated $K$--factors are displayed in the right--most column. 
The total rate central values are folded with the theory uncertainty estimates
from scale variations (first quote) and PDFs (second quote).
The LHC center--of mass energy is  $\sqrt{S} = 14$ TeV. The 2HDM parameters
are fixed to benchmark B1 in Table ~3.}
\label{tab:B1table}
\end{table} 

\begin{table} [tb!]
\begin{center}
    \begin{tabular}{ | c || r | r | r |}
   \multicolumn{4}{l}{\textbf{Benchmark B1}} \\      \hline \hline  
  \textbf{$q\bar{q}$--channels}   & LO & NLO  & $K$--factor\\ \hline
    $\hzero\Azero$ & 0.0181  $^{+3.7+1.7\%}_{-3.6-1.7\%}$ & 0.0232 $^{+2.1+2.3\%}_{-1.8-1.8\%}$ & 1.29 \\ \hline
    $\Hzero\Azero$ & 1.53  $^{+5.2+1.9\%}_{-4.8-1.9\%}$ &  1.95  $^{+2.2+2.4\%}_{-2.0-1.9\%}$ & 1.27  \\ \hline
    $\PHiggs^+\PHiggs^-$ & 0.814  $^{+6.1+2.1\%}_{-5.6-2.1\%}$  & 1.02 $^{+2.3+2.6\%}_{-2.2-1.9\%}$ & 1.25 \\  \hline  \hline 
   \end{tabular}
\end{center}
\caption{Total Higgs pair cross sections via quark--antiquark annihilation
$pp(q\bar{q}) \to h_i\,h_j$ (in fb)
for the mixed \CP--even/\CP--odd and charged Higgs pair combinations within the 2HDM, in the same
setup as Table~5. The 2HDM parameters
are fixed to benchmark B1 in Table~3.}
\label{tab:B1qqtable}
\end{table}

Before we discuss the characteristic model--dependent features, 
some general comments are in order. First of all, 
we find large NLO corrections, with typical $K$--factors
in the range of $1.5 - 1.7$ for the $gg$--initiated channels. 
These sizable QCD effects are primarily due to the
NLO real emission.  This is a well--known fact for processes
with color--singlet final states, where there is plenty of phase
space to accommodate the radiation of initial--state light partons \cite{Ohnemus:1990za,Ohnemus:1991gb,Ohnemus:1991kk,Dolan:2012rv}.
Notice that these $K$--factors do not
change significantly as we compare the 
different $gg$--initiated
di--Higgs final--states. Likewise, they do not depend on the chosen benchmark and 
they are quantitatively similar to the SM prediction. We note here for completeness that the corresponding predictions for the SM 
give a LO cross section of 23.0~fb, a NLO ``loop-improved" result of 34.9~fb and a NLO ``Born-improved'' one of 38.9~fb \cite{Frederix:2014hta} and therefore a $K$--factor of 1.52.
These traits can be again explained by the structure of the QCD corrections (cf. Section~\ref{sec:nlo}),
which is common to all Higgs pair channels and unlinked to the underlying 2HDM dynamics. 
Actually, all genuine 2HDM imprints (viz. the modified couplings and the heavy resonances) modify
the LO and NLO predictions in exactly the same way, leaving
the $K$--factor values unaltered. 

\smallskip{}
The reduction of the 
theoretical uncertantities is manifest in the
smaller scale variations when we compare the total rates at LO and NLO accuracy. 
For gluon fusion, we find typical scale uncertainties spanning
$\Delta \sigma/\sigma \equiv [\sigma(\mu^0/2)-\sigma(2\mu^0)]/\sigma \sim 30-40\%$ at LO, while they shrink
down to $\Delta \sigma/\sigma \sim 15-20\%$ at NLO. 
The PDF uncertainties lie at the per--cent level and
increase with heavier Higgs masses. This behavior can be attributed 
to the uncertainty
in the gluon parton density, which increases with the Bjorken variable $x$
in the kinematically relevant regions for these processes. 
On the other hand, the $q\bar{q}$--initiated subprocesses exhibit an overall milder scale uncertainty,
in line with the fact that they do not depend on $\alpha_s$ at LO. The corresponding $K$--factors
are also smaller in these cases (cf. Tables~\ref{tab:B1qqtable}, \ref{tab:B2qqtable} and ~\ref{tab:B3qqtable}) and
remain in the range $K\sim 1.2-1.3$, as expected for Drell--Yan--like
processes. 
Notice also that, unlike the scale uncertainties, the PDF uncertainties 
slightly grow from the LO
to the NLO predictions. This is explained by the additional initial--state
partons which become active at NLO. The differences between the ``loop--improved'' and ``born--improved'' results are
typically smaller than 10\% and therefore lie within the theoretical uncertainties.
Another aspect to mention is
the relative size of the (tree--level) $q\bar{q}$--initiated
channels versus the  
(loop--induced) $gg$--fusion mechanism. This varies significantly with the 
different Higgs pair combinations. For instance, while  
gluon fusion  
prevails for $\hzero\Azero$, in the case of
$\Hzero\Azero$ we find that the bulk contribution is $q\bar{q}$ induced. 
This difference can be traced back to the coupling $g_{\hzero\Azero\PZ^0}$ (resp. $g_{\Hzero\Azero\PZ^0}$), 
which is proportional to $\cos(\beta-\alpha)$ (resp. sin$(\beta-\alpha)$) and therefore vanishes (resp.
maximizes) in the decoupling limit $\xi = \cos(\beta-\alpha) \to 0$.

\begin{table}[tb!]
\begin{center}
    \begin{tabular}{ | c || r | r | r | r |}
   \multicolumn{5}{l}{\textbf{Benchmark B2}} \\ \hline \hline
     \textbf{$gg$--channels} & LO &NLO--loop improved &NLO--Born improved  & $K$--factor\\ \hline \hline
    $\hzero\hzero$ & 85.1$^{+33.5+2.3\%}_{-23.5-2.5\%}$ & 135$^{+15.9+2.3\%}_{-14.0-2.8 \%}$ & 147$^{+18.4+2.3\%}_{-15.3-2.8 \%}$ & 1.59 \\ \hline
    $\hzero\Hzero$ & 0.849$^{+36.9+3.3\%}_{-25.2-3.4\%}$ & 1.42$^{+14.5+3.4\%}_{-14.0-4.3 \%}$  & 1.51$^{+16.5+3.5\%}_{-15.0-4.3 \%}$  & 1.67 \\ \hline
    $\Hzero\Hzero$ & 0.00763$^{+40.5+5.1\%}_{-26.9-5.0\%}$  & 0.0126$^{+15.8+5.9\%}_{-15.3-6.9 \%}$  & 0.0132$^{+17.4+5.9\%}_{-16.1-6.7 \%}$  & 1.65  \\ \hline 
    $\hzero\Azero$ & 0.607$^{+36.7+3.2\%}_{-25.1-3.3\%}$  & 0.986$^{+16.9+3.3\%}_{-15.1-4.1 \%}$ & 1.018$^{+17.9+3.3\%}_{-15.6-4.1 \%}$  & 1.62 \\ \hline 
    $\Hzero\Azero$ & 0.0051$^{+40.2+4.8\%}_{-26.8-4.8\%}$  & 0.0078$^{+13.9+5.8\%}_{-14.4-6.8 \%}$ & 0.0088$^{+17.5+5.6\%}_{-16.1-6.5 \%}$  & 1.53 \\ \hline 
    $\Azero\Azero$ & 0.0159 $^{+40.2+4.9\%}_{-26.8-4.9\%}$& 0.0246$^{+13.4+5.7\%}_{-14.1-6.7 \%}$ &  0.0272$^{+16.8+5.6\%}_{-15.7-6.5 \%}$ & 1.54  \\ \hline
    $\PHiggs^+\PHiggs^-$ & 0.0243$^{+40.0+4.7\%}_{-26.7-4.8 \%}$ & 0.0393$^{+15.2+5.6\%}_{-15.0-6.5 \%}$ &  0.0424$^{+17.6+5.5\%}_{-16.1-6.3 \%}$ & 1.62 \\ \hline   \hline
   \end{tabular}
\end{center}
\caption{Total Higgs pair cross sections via gluon fusion $pp(gg) \to h_i\,h_j$ (in fb)
for all seven final--state combinations within the 2HDM, in the same setup as Table~5. The 2HDM parameters
are fixed to benchmark B2 in Table~3.}
\label{tab:B2table}
\end{table} 

\begin{table} [tb!]
\begin{center}    
\begin{tabular}{ | c || r | r | r |}
   \multicolumn{4}{l}{\textbf{Benchmark B2}} \\      \hline \hline  
    \textbf{$q\bar{q}$--channels}  & LO & NLO  & $K$--factor\\ \hline
    $\hzero\Azero$ & 5.59$\cdot$10$^{-3}$ $^{+6.2+2.1\%}_{-5.6-2.0\%}$ & 7.08$\cdot$10$^{-3}$ $^{+2.2+2.6\%}_{-2.2-1.9\%}$ & 1.27 \\ \hline
    $\Hzero\Azero$ & 8.26$\cdot$10$^{-2}$ $^{+9.1+3.0\%}_{-7.9-2.6\%}$ &  0.100 $^{+2.7+3.4\%}_{-2.9-2.2\%}$ & 1.21 \\ \hline
    $\PHiggs^+\PHiggs^-$ & 0.117$^{+8.6+2.9\%}_{-7.5-2.7\%}$ & 0.142 $^{+2.6+3.3\%}_{-2.8-2.2\%}$ & 1.22 \\  \hline \hline  
   \end{tabular}
\end{center}
\caption{Total Higgs pair cross sections via quark--antiquark annihilation
$pp(q\bar{q}) \to h_i\,h_j$ (in fb)
for the mixed \CP--even/\CP--odd and charged Higgs pair combinations within the 2HDM, in the same
setup as Table~5. The 2HDM parameters
are fixed to benchmark B2 in Table~3.}
\label{tab:B2qqtable}
\end{table} 

\medskip{}
While the $K$--factors barely depend on the specific 2HDM scenario, the total
rates critically rely on the chosen benchmark. In the light Higgs pair case,
these can vary from $\sigma^{\text{NLO}} \sim 30$ fb up 
to $\sigma^{\text{NLO}} \sim 2$ pb. 
The latter pb--level rates are
roughly two orders of magnitude above the SM expectations, and  
are linked to the resonant heavy Higgs contribution $gg \to \Hzero \to \hzero\hzero$. 
This trait is common to all scenarios in which the cascade decay $\Hzero \to \hzero\hzero$
is kinematically accessible ($m_{\Hzero} > 2 m_{\hzero}$) 
and reflects the fact that the resonant
production effectively involves one power of $G_F$ less compared to the continuum pair production. 
Such enhancements are tamed if the position of the resonance is shifted away from the di--Higgs threshold 
$2m_{\hzero}$ and
vanish  
 consistently in the SM limit of $\xi \to 0$ and $ m_{\Hzero} \gg m_{\hzero}$.  
The fact that the resonant contribution to $\sigma(\hzero\hzero)$ decreases with increasing $m_{\Hzero}$ values is explained by i) the
larger phase space required to produce the intermediate heavy state $\Hzero$; and ii) the correspondingly lower gluon luminosity, 
which is probed at larger $x$--values, the more massive the heavy field becomes.
It is also remarkable that these resonant situations
are almost insensitive to the actual value of
the triple Higgs self--couplings. This is due to the fact that i) the dependence 
of the resonant production $gg \to \Hzero$ on the coupling $g_{\Hzero\hzero\hzero}$  is cancelled
by the dominant decay mode $\Hzero \to \hzero\hzero$ and ii)  
the light Higgs--mediated diagrams $gg \to \hzero^* \to \hzero\hzero$ 
are subdominant with respect to the resonant part, in such a way that
the dependence on the coupling $g_{\hzero\hzero\hzero}$ is overshadowed.

\medskip{}
A number of model--specific fingerprints can be unraveled by comparing 
the total rate predictions for the different 2HDM benchmarks. 
These can be ultimately traced back to the characteristic
coupling patterns and the extra Higgs boson masses in each scenario, given in Tables ~\ref{tab:benchmarks} 
and ~\ref{tab:couplings}.
For the resonant scenarios, B1 and B2 , the 
on--shell production of the heavy Higgs field $\Hzero$ is responsible
for the total rate enhancement in $\sigma(\hzero\hzero)$, by a factor $70$ (resp. 4) 
above the SM. The difference of one order
of magnitude between the two scenarios is linked to the fact
that, while in B1 the heavy Higgs mass is relatively
low and lies right above the di--Higgs threshold $m_{\Hzero} \simeq 2m_{\hzero}$, 
in scenario B2 the $\Hzero$ field is substantially heavier. The total $\hzero\hzero$
rates in the first case thus benefit from the larger on--shell
single $\Hzero$ rates; as well as from the overly dominant decay mode $\Hzero \to \hzero\hzero$.
For scenarios B6 and B7 in Table~\ref{tab:b5b6b7} we find 
milder ($\sim$ 30\%) resonant enhancements of the $\hzero\hzero$ cross section as
compared to B1, 
even though the $\Hzero$ mass is also quite low in these cases. This is because
the $\Hzero$ contribution through
$gg \to \Hzero \to \hzero\hzero$ is suppressed by the reduced heavy Higgs top Yukawa $g_t^{\Hzero}$. 
Finally, for the non--resonant scenarios (viz. B3, B4 and B5) our predictions for $\sigma(\hzero\hzero)$
fall down to values close, or even slightly below the SM. 
This is explained naturally by the SM--like coupling pattern
of B3 and B5 (cf. Table~\ref{tab:couplings}). For B4, the reduced rate
is a consequence of the stronger destructive interference between i) the triangle--mediated
contributions $gg \to \hzero^* \to \hzero\hzero$ and  
$gg \to \Hzero^* \to \hzero\hzero$, which are enhanced by the larger
trilinear couplings; and ii) the box--mediated
diagrams.
Barring possible  
on--shell resonances, we conclude that significant deviations
from the SM Higgs pair predictions are not possible rate--wise. This is 
in agreement with the findings reported in Ref.~\cite{Baglio:2014nea}.
Similarly, we see that the distinctive model--specific
features have very little impact on the total rates and on the size of the QCD corrections.

\begin{table} [tb!]
\begin{center}
    \begin{tabular}{ | c || r | r | r | r |}
   \multicolumn{5}{l}{\textbf{Benchmark B3}} \\ \hline \hline
     \textbf{$gg$--channels} & LO &NLO--loop improved &NLO--Born improved & $K$--factor \\ \hline
    $\hzero\hzero$ & 22.1$^{+32.4+2.0\%}_{-22.9-2.3\%}$ & 33.9$^{+15.2+2.0\%}_{-13.5-2.4\%}$ & 37.5$^{+18.3+2.0\%}_{-15.1-2.4\%}$  & 1.53 \\ \hline
    $\hzero\Hzero$ & 11.1$^{+32.5+2.1\%}_{-23.0-2.4\%}$ & 17.1$^{+15.1+2.1\%}_{-13.4-2.5\%}$ & 19.0$^{+18.5+2.0\%}_{-15.1-2.5\%}$  & 1.53 \\ \hline
    $\Hzero\Hzero$ &  0.702$^{+33.2+2.2\%}_{-23.3-2.5\%}$ & 1.08$^{+14.3+2.2\%}_{-13.1-2.7\%}$  & 1.22$^{+17.7+2.2\%}_{-14.9-2.7\%}$  & 1.54 \\ \hline
    $\hzero\Azero$ & 6.53$^{+33.5+2.3\%}_{-23.5-2.5\%}$  & 11.1$^{+18.1+2.2\%}_{-15.2-2.7\%}$  & 12.8$^{+18.3+2.2\%}_{-15.3-2.7\%}$  & 1.70 \\ \hline 
    $\Hzero\Azero$ & 2.51$^{+34.6+2.6\%}_{-24.0-2.7\%}$  & 4.08$^{+16.1+2.6\%}_{-14.4-3.2\%}$  &  4.39$^{+18.3+2.6\%}_{-15.5-3.2\%}$   & 1.62 \\ \hline     
    $\Azero\Azero$ & 0.485$^{+36.0+3.0\%}_{-24.7-3.1\%}$   & 0.787$^{+15.7+3.1\%}_{-14.4-3.9\%}$ & 0.840$^{+17.7+3.1\%}_{-15.4-3.9\%}$   & 1.62 \\ \hline
    $\PHiggs^+\PHiggs^-$ & 0.840$^{+35.9+3.0\%}_{-24.7-3.1\%}$  &1.40$^{+15.7+3.1\%}_{-14.4-3.9\%}$   & 1.50$^{+17.8+3.0\%}_{-15.4-3.8\%}$    & 1.67 \\ \hline \hline  
   \end{tabular}
\end{center}
\caption{Total Higgs pair cross sections via gluon fusion $pp(gg) \to h_i\,h_j$ (in fb)
for all seven final--state combinations within the 2HDM, in the same setup as Table~5. The 2HDM parameters
are fixed to benchmark B3 in Table~3.}
\label{tab:B3table}
\end{table} 

\begin{table} [tb!]
\begin{center}
\begin{tabular}{ | c || r | r | r |}
   \multicolumn{4}{l}{\textbf{Benchmark B3}} \\      \hline \hline  
    \textbf{$q\bar{q}$--channels}  & LO & NLO  & $K$--factor\\ \hline
    $\hzero\Azero$ & 1.96$\cdot$10$^{-3}$ $^{+2.6+1.5\%}_{-2.7-1.6\%}$ & 2.55$\cdot$10$^{-3}$ $^{+1.9+2.2\%}_{-1.6-1.7\%}$ & 1.30 \\ \hline
    $\Hzero\Azero$ & 5.02 $^{+3.4+1.6\%}_{-3.4-1.7\%}$ &  6.50 $^{+2.1+2.2\%}_{-1.8-1.7\%}$ & 1.29 \\ \hline
    $\PHiggs^+\PHiggs^-$ & 2.17 $^{+4.7+1.8\%}_{-4.4-1.9\%}$ & 2.76 $^{+2.1+2.4\%}_{-1.9-1.8\%}$ & 1.27 \\  \hline  \hline 
   \end{tabular}
\end{center}
\caption{Total Higgs pair cross sections via quark--antiquark annihilation
$pp(q\bar{q}) \to h_i\,h_j$ (in fb)
for the mixed \CP--even/\CP--odd and charged Higgs pair combinations within the 2HDM, in the same
setup as Table~5. The 2HDM parameters
are fixed to benchmark B3 in Table~3.}
\label{tab:B3qqtable}
\end{table}

\begin{table} [tb!]
\begin{center}
    \begin{tabular}{ | c || r | r | r | r|}
    \multicolumn{5}{l}{\textbf{Benchmarks B4 -- B7  }} \\      \hline \hline 
    \textbf{$pp(gg) \to \hzero\hzero$}  & LO & NLO--loop improved  &  NLO--Born improved & $K$--factor\\ \hline \hline
    \textbf{B4} & 18.0$^{+32.0+1.9\%}_{-22.7-2.3\%}$ & 28.7$^{+16.6+2.0\%}_{-14.1-2.4\%}$ & 30.5$^{+18.4+1.9\%}_{-15.0-2.3\%}$ & 1.59  \\ \hline
    \textbf{B5} & 23.4$^{+32.3+2.0\%}_{-22.9-2.3\%}$ & 35.3 $^{+15.1+2.0\%}_{-13.4-2.4\%}$ & 39.4$^{18.4+2.0\%}_{-15.1-2.4\%}$ & 1.50  \\ \hline
    \textbf{B6} & 30.3  $^{+32.4+2.0\%}_{-22.9-2.3\%}$ &  45.5 $^{+14.5+2.0\%}_{-13.1-2.5\%}$ & 51.8$^{+18.4+2.0\%}_{-15.1-2.4\%}$ & 1.50 \\ \hline
    \textbf{B7} & 29.1  $^{+32.6+2.1\%}_{-23.0-2.4\%}$  & 44.9 $^{+15.4+2.1\%}_{-13.6-2.5\%}$ & 49.5$^{+18.3+2.0\%}_{-15.1-2.5\%}$& 1.54 \\  \hline \hline  
   \end{tabular}
\end{center}
\caption{Total light Higgs pair cross sections via gluon fusion $pp(gg) \to \hzero\hzero$ (in fb)
within the 2HDM, in the same setup as Table~5. The 2HDM parameters
are fixed to benchmarks B4 -- B7 in Table~3.}
\label{tab:b5b6b7}
\end{table}

\smallskip{}
Before closing this discussion, let us devote one last word to the 
heavier di--Higgs combinations, shown in  
Tables~\ref{tab:B1table}, \ref{tab:B2table} and ~\ref{tab:B3table} for
the B1, B2 and B3 scenarios respectively. The reported total rates
span a wide range from $\sigma\sim \mathcal{O}(10^{-2})$ fb to
$\sigma\sim \mathcal{O}(10)$ fb and lie in all cases below
the light di--Higgs pair predictions. The reason is twofold:
i) the relative phase space suppression and ii) the lower initial--parton luminosities
involved in the production of these heavier states. 
The cases in which different rates are obtained
for different Higgs pair combinations of similar masses
(e.g. for $\hzero\Hzero$
and $\Hzero\Hzero$ in benchmark B3) can be understood by considering the size of the
Higgs couplings in the given scenario.

\subsection{Differential distributions}
\label{sec:distributions}

As we have seen so far, the most apparent
2HDM imprints rate--wise are of resonant nature. These 
appear when the on--shell subprocess $gg \to\Hzero \to \hzero\hzero$
is kinematically available and adds to the continuum production. 
This possibility crucially depends on
the heavy Higgs spectrum and is almost insensitive 
to the distinctive coupling patterns of the 2HDM. The mere observation
of an enhancement in the total rate would thus not be undisputed
evidence of an underlying 2HDM structure.

\begin{figure}[tb!]
 \begin{minipage}[t]{0.5\linewidth}
\centering
\includegraphics[trim=1cm 0 0 0,scale=0.56]{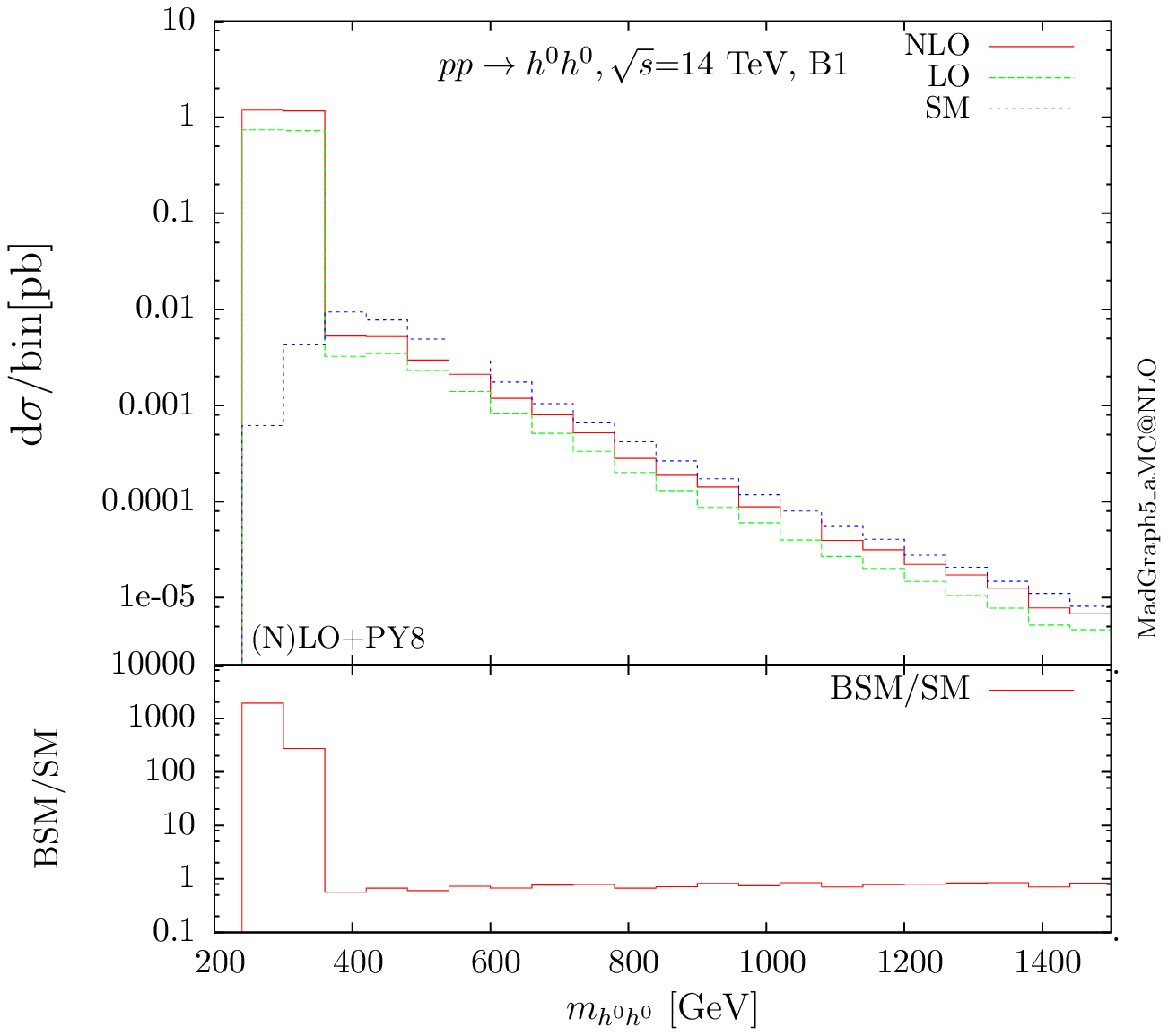}
\label{h1h1_H1}
\end{minipage}
\hspace{0.5cm}
 \begin{minipage}[t]{0.5\linewidth}
 \centering
 \includegraphics[trim=1.5cm 0 0 0,scale=0.56]{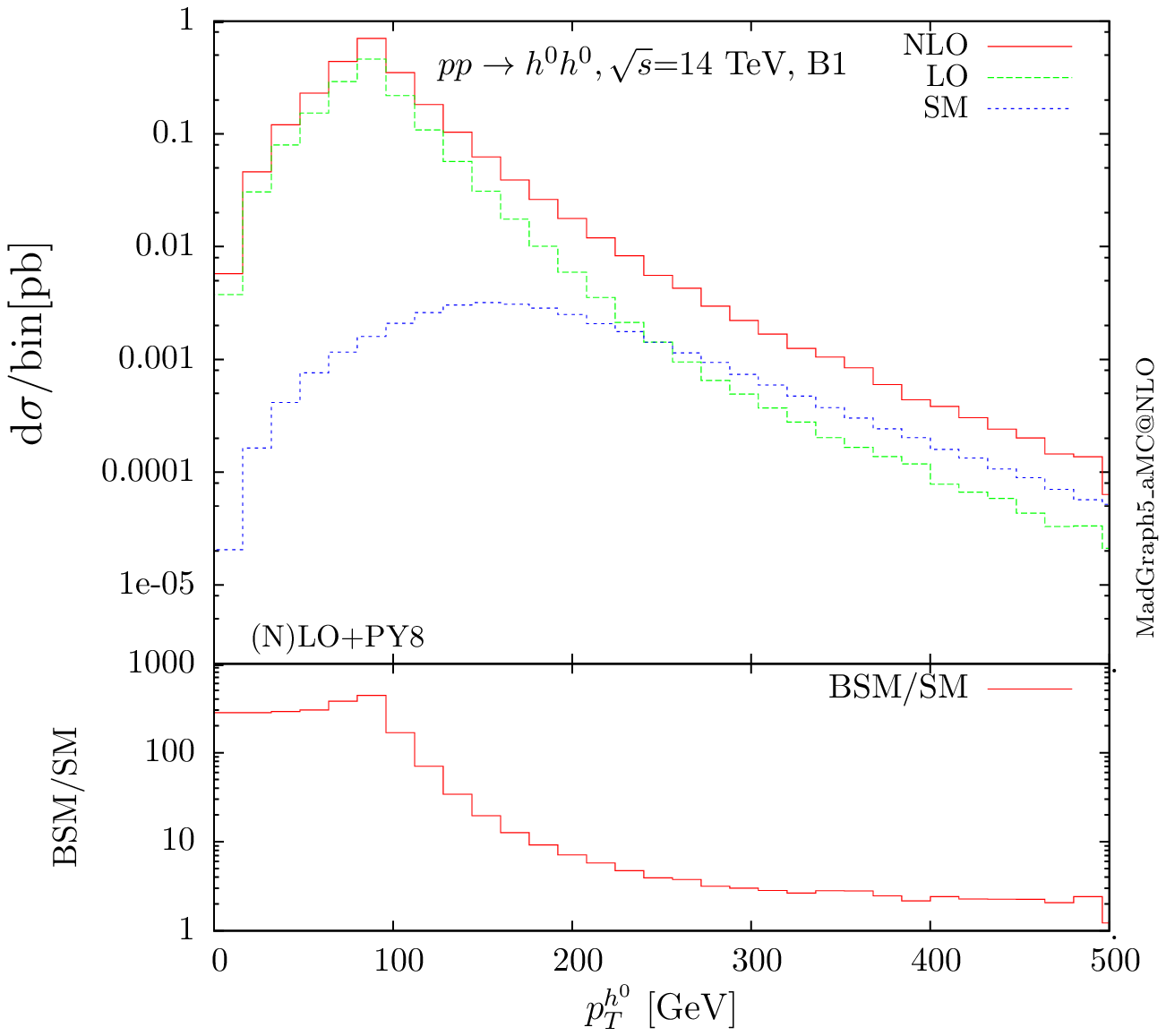}
 \end{minipage}
\caption{\label{fig:b1} Light Higgs pair differential rates as a function 
of a) the di--Higgs invariant mass 
$m_{\hzero\hzero}$ (left panels); 
and b) the hardest Higgs transverse momentum $p_T^{\hzero}$ (right panels). 
We separately show 
the results at LO+PS and NLO+PS accuracy in QCD, where the latter correspond to the
``loop--improved'' approach. The NLO+PS prediction for the SM is overlayed for comparison. In the lower subpannels
we display the bin--by--bin ratio of the 2HDM prediction at NLO+PS over the corresponding SM result. The LHC center--of--mass energy
is $\sqrt{S} = 14$ TeV. The 2HDM parameters are fixed to benchmark B1.} 
\end{figure}

\begin{figure}[tb!]
 \begin{minipage}[t]{0.5\linewidth}
\centering
\includegraphics[trim=1cm 0 0 0,scale=0.56]{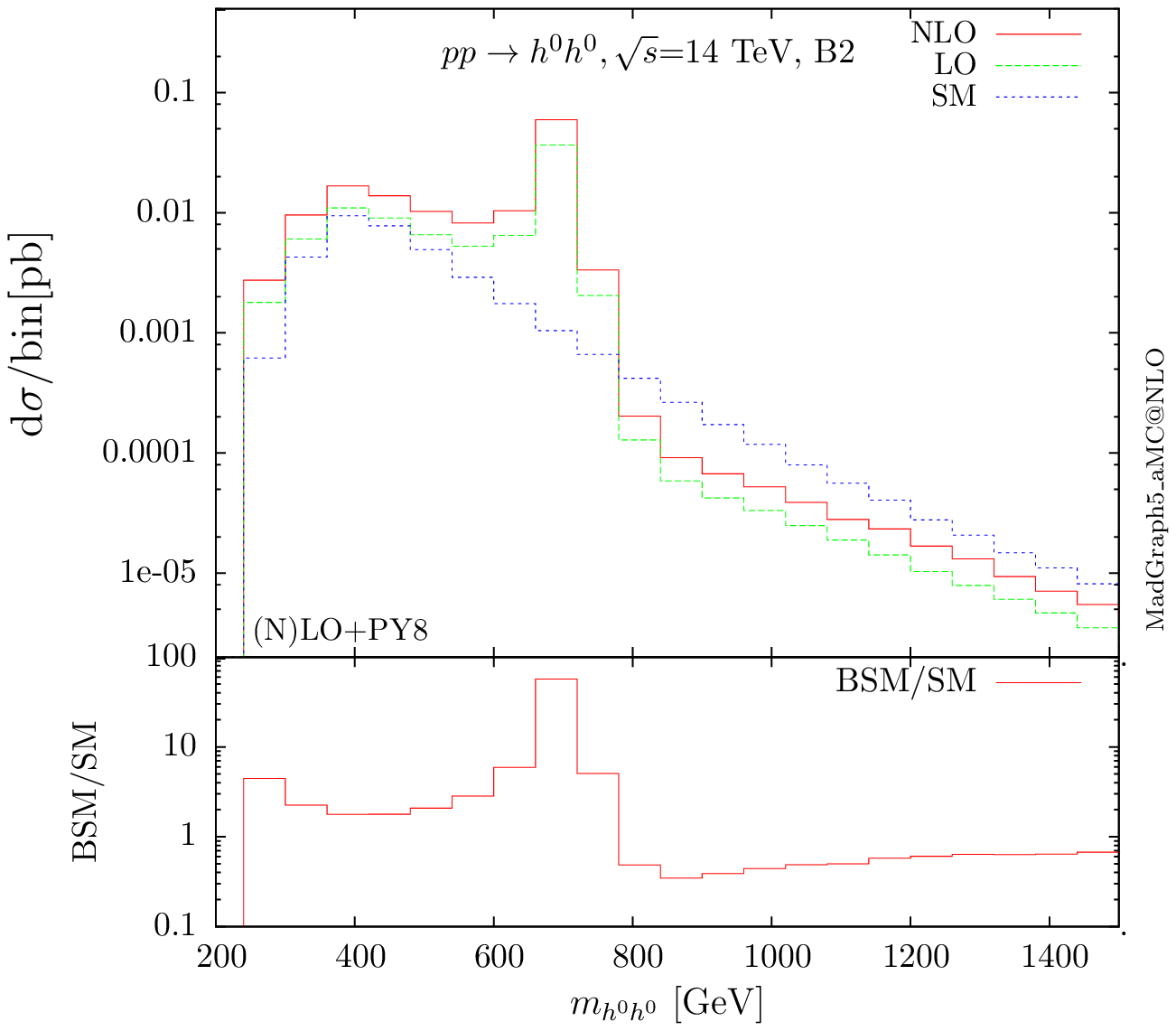}
\end{minipage}
\hspace{0.5cm}
 \begin{minipage}[t]{0.5\linewidth}
 \centering
 \includegraphics[trim=1.5cm 0 0 0,scale=0.56]{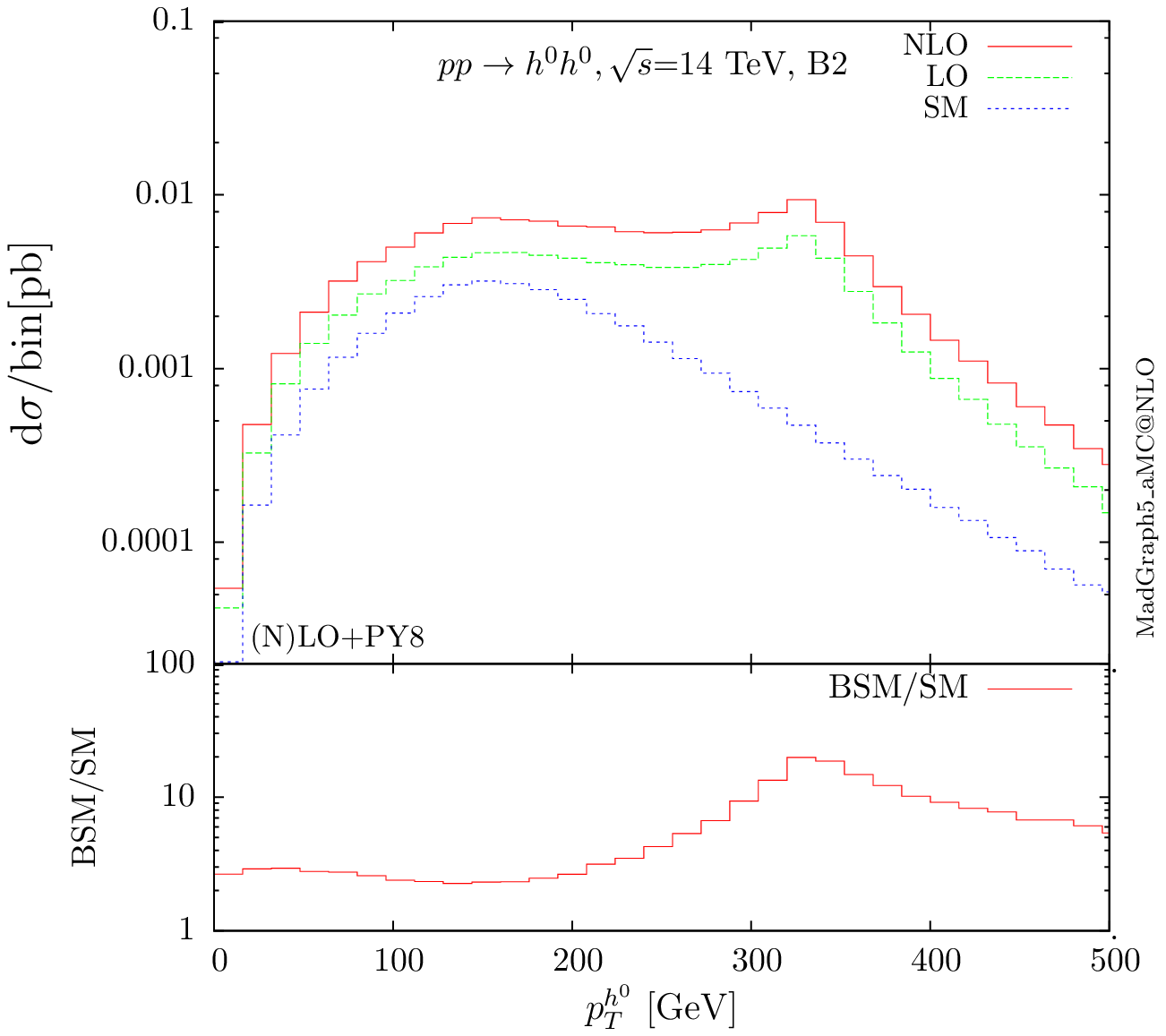}
\label{lptmet}
 \end{minipage}
 \caption{\label{fig:b2}  Light Higgs pair differential rates as a function 
of a) the di--Higgs invariant mass 
$m_{\hzero\hzero}$ (left panels); 
and b) the hardest Higgs transverse momentum $p^{\hzero}_T$ (right panels), in the same setup as for Figure~4. The 2HDM parameters are fixed as in benchmark B2.} 
\end{figure}

\begin{figure}[tb!]
 \begin{minipage}[b]{0.5\linewidth}
\centering
\includegraphics[trim=1cm 0 0 0,scale=0.56]{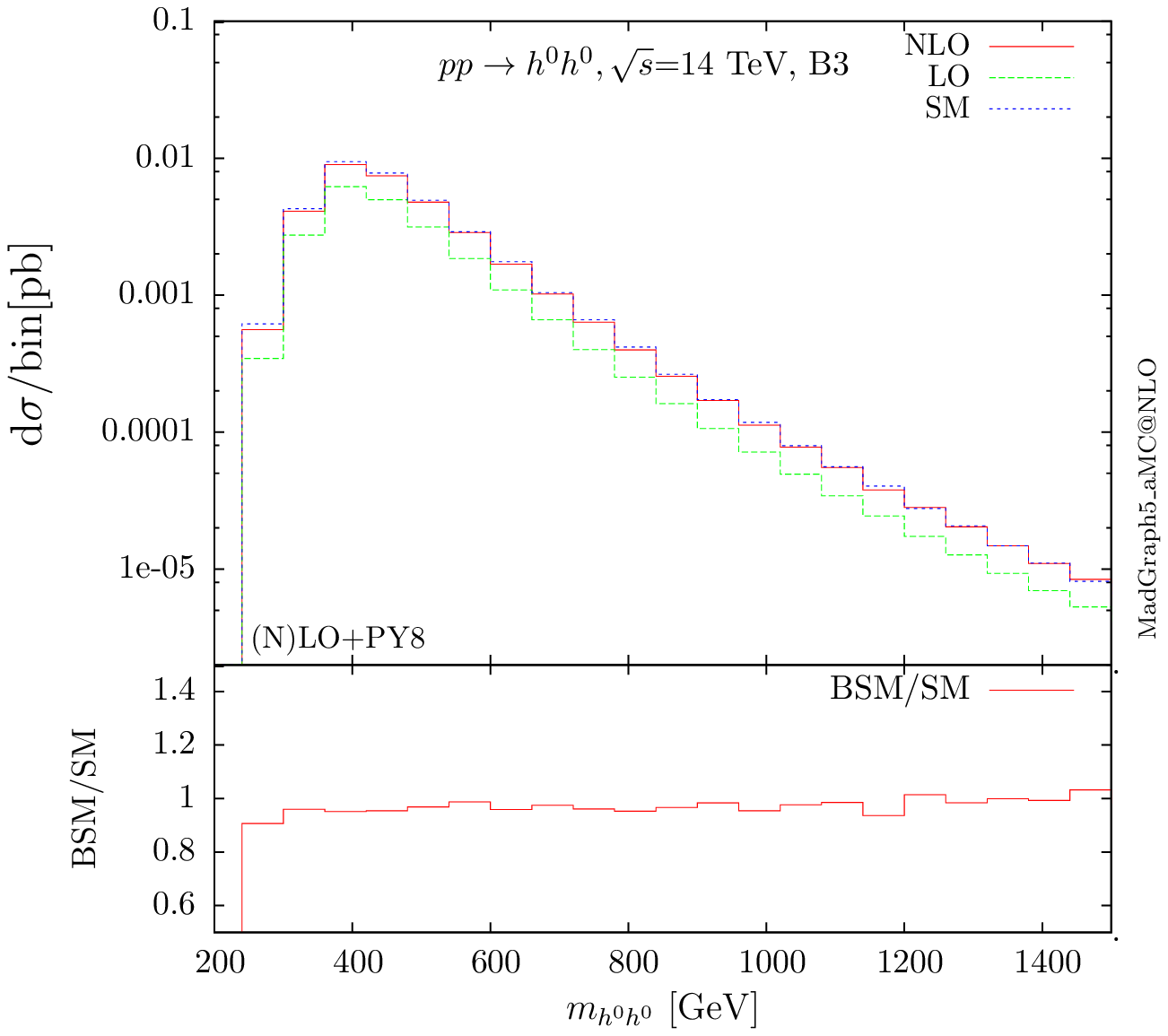}
\end{minipage}
\hspace{0.5cm}
 \begin{minipage}[b]{0.5\linewidth}
 \centering
 \includegraphics[trim=1.5cm 0 0 0,scale=0.56]{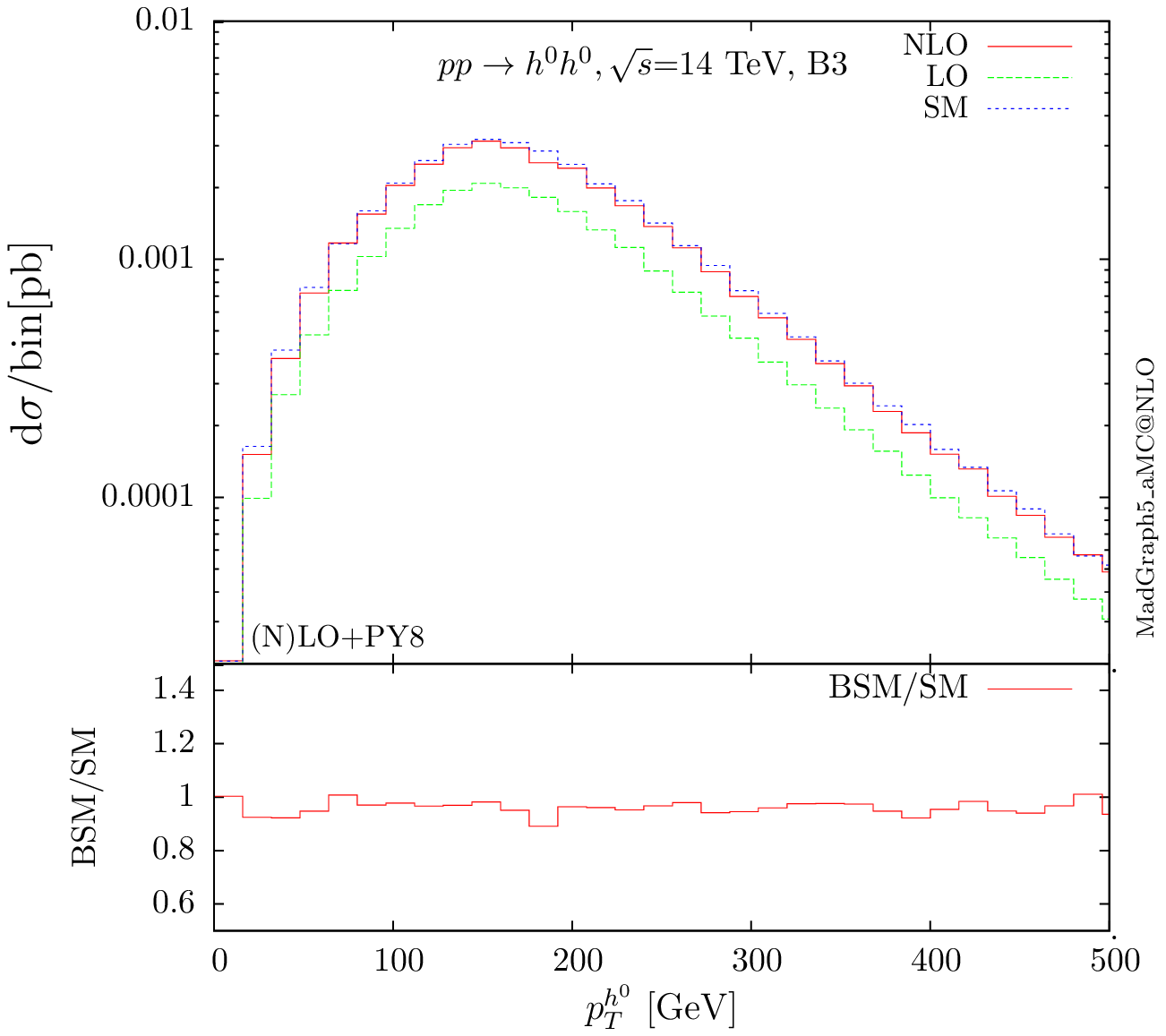}
 \end{minipage}
 \caption{\label{fig:b3} Light Higgs pair differential rates as a function 
of a) the di--Higgs invariant mass 
$m_{\hzero\hzero}$ (left panels); 
and b) the hardest Higgs transverse momentum $p^{\hzero}_T$ (right panels), in the same setup as for Figure~4. The 2HDM parameters are fixed as in benchmark B3.} 
\end{figure}

\begin{figure}[tb!]
 \begin{minipage}[b]{0.5\linewidth}
\centering
\includegraphics[trim=1cm 0 0 0,scale=0.56]{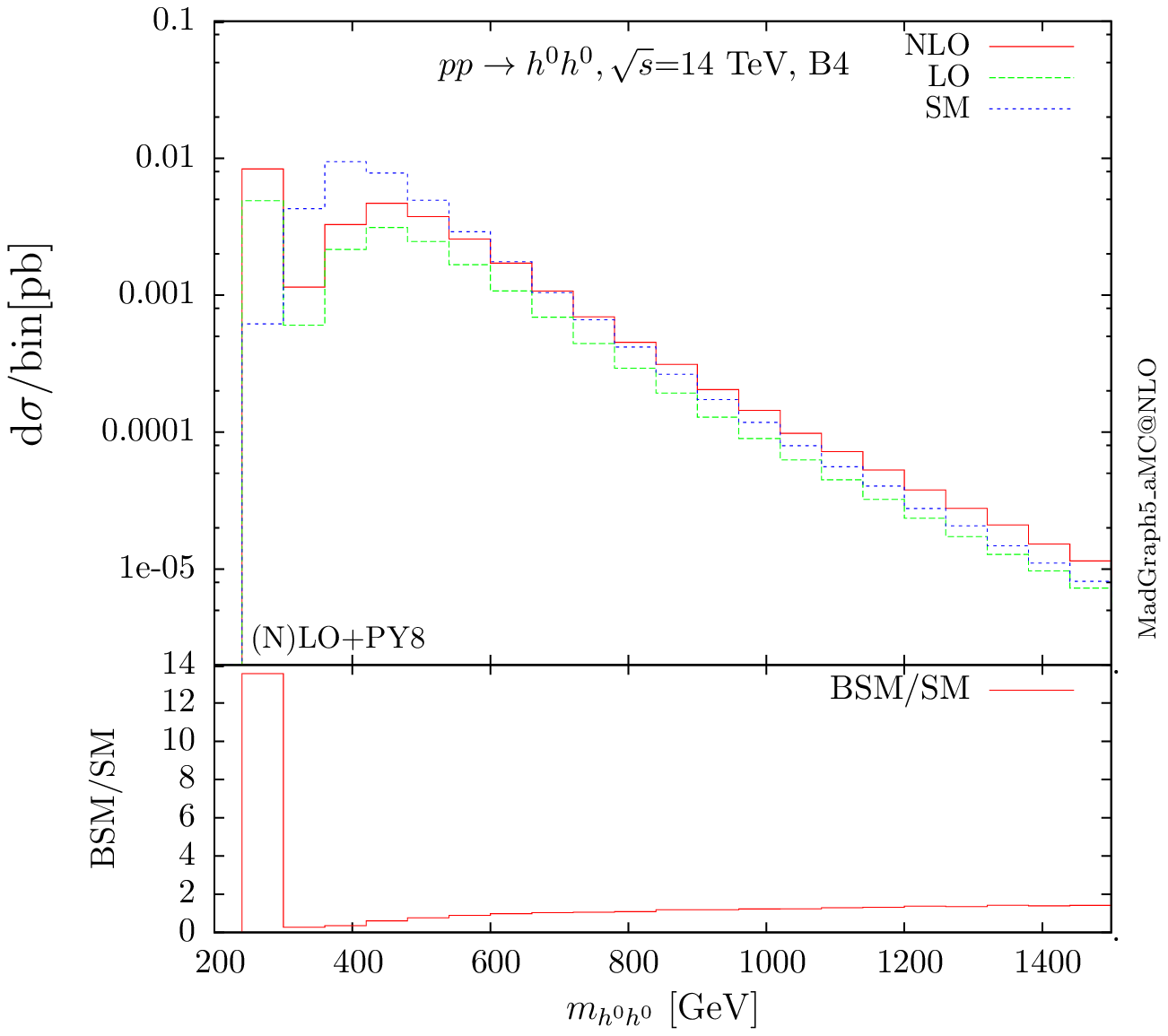}
\end{minipage}
\hspace{0.5cm}
 \begin{minipage}[b]{0.5\linewidth}
 \centering
 \includegraphics[trim=1.5cm 0 0 0,scale=0.56]{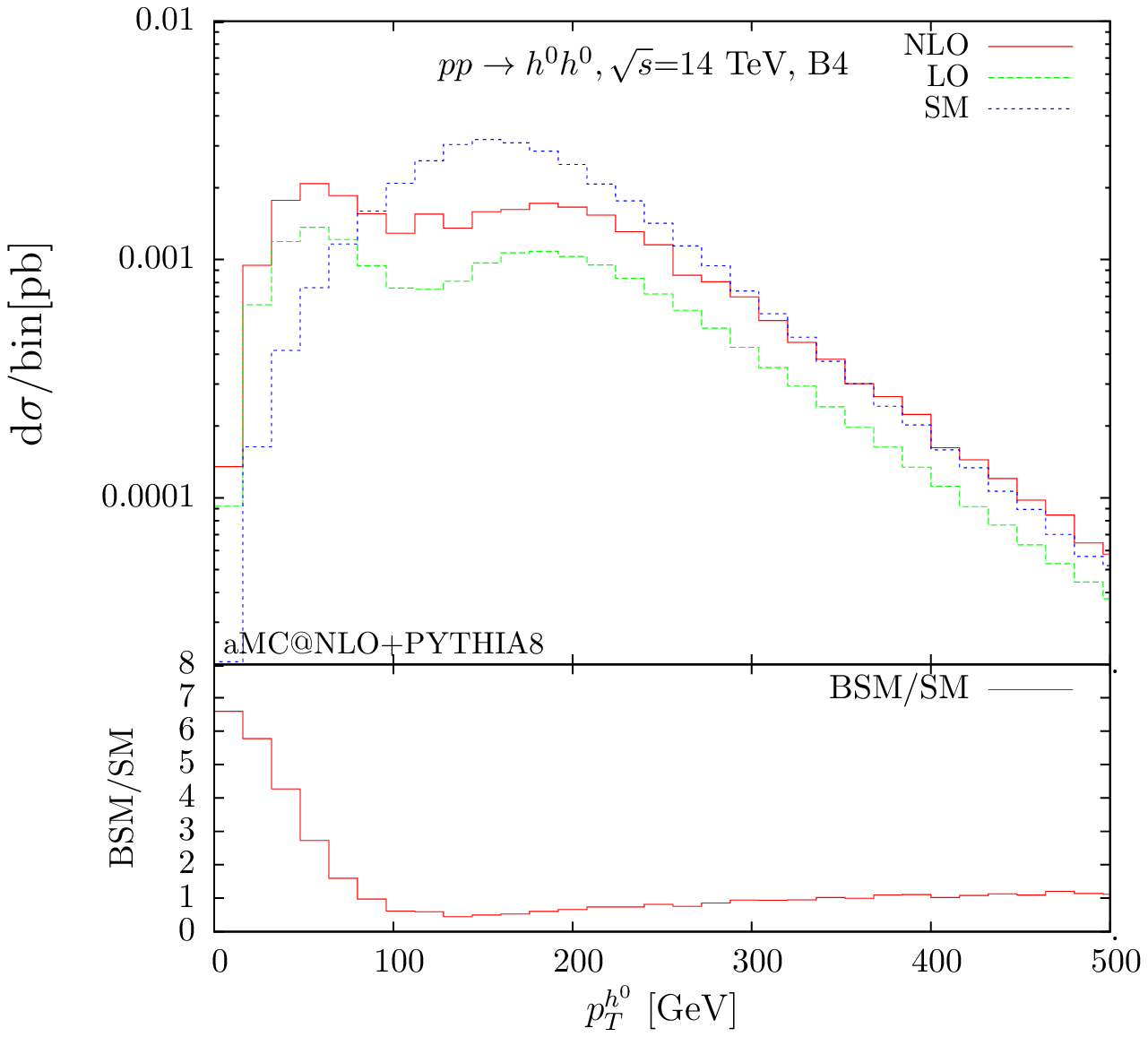}
 \end{minipage}
 \caption{\label{fig:b4} Light Higgs pair differential rates as a function 
of a) the di--Higgs invariant mass 
$m_{\hzero\hzero}$ (left panels); 
and b) the hardest Higgs transverse momentum $p^{\hzero}_T$ (right panels), in the same setup as for Figure~4. The 2HDM parameters are fixed as in benchmark B4.} 
\end{figure}

\begin{figure}[tb!]
\centering
\includegraphics[trim=1cm 0 0 0,scale=0.65]{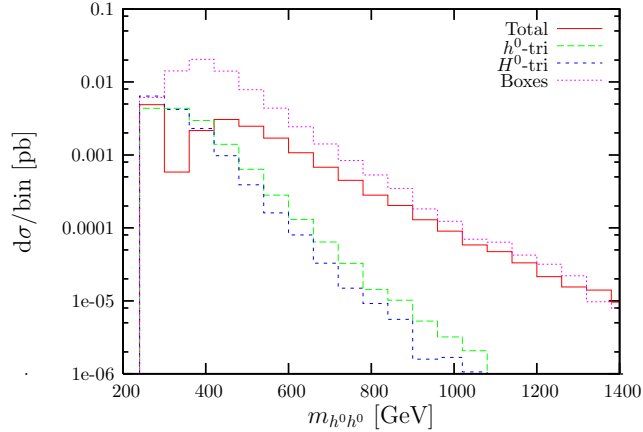}
\caption{\label{fig:mhhb4} Individual contributions to the light 
Higgs pair differential rates as a function 
of the di--Higgs invariant mass 
$m_{\hzero\hzero}$. The results are shown separately for i) the
light Higgs--mediated triangles (long--dashed, green), ii) the heavy Higgs--mediated triangles (short--dashed, blue);
 iii) the box topologies (dotted, magenta); and iv) the combined contribution. All histograms
 are computed to LO+PS accuracy. The 2HDM parameters are fixed as in benchmark B4.} 
\end{figure}

\begin{figure}[tb!]
 \begin{minipage}[b]{0.5\linewidth}
\centering
\includegraphics[trim=1cm 0 0 0,scale=0.56]{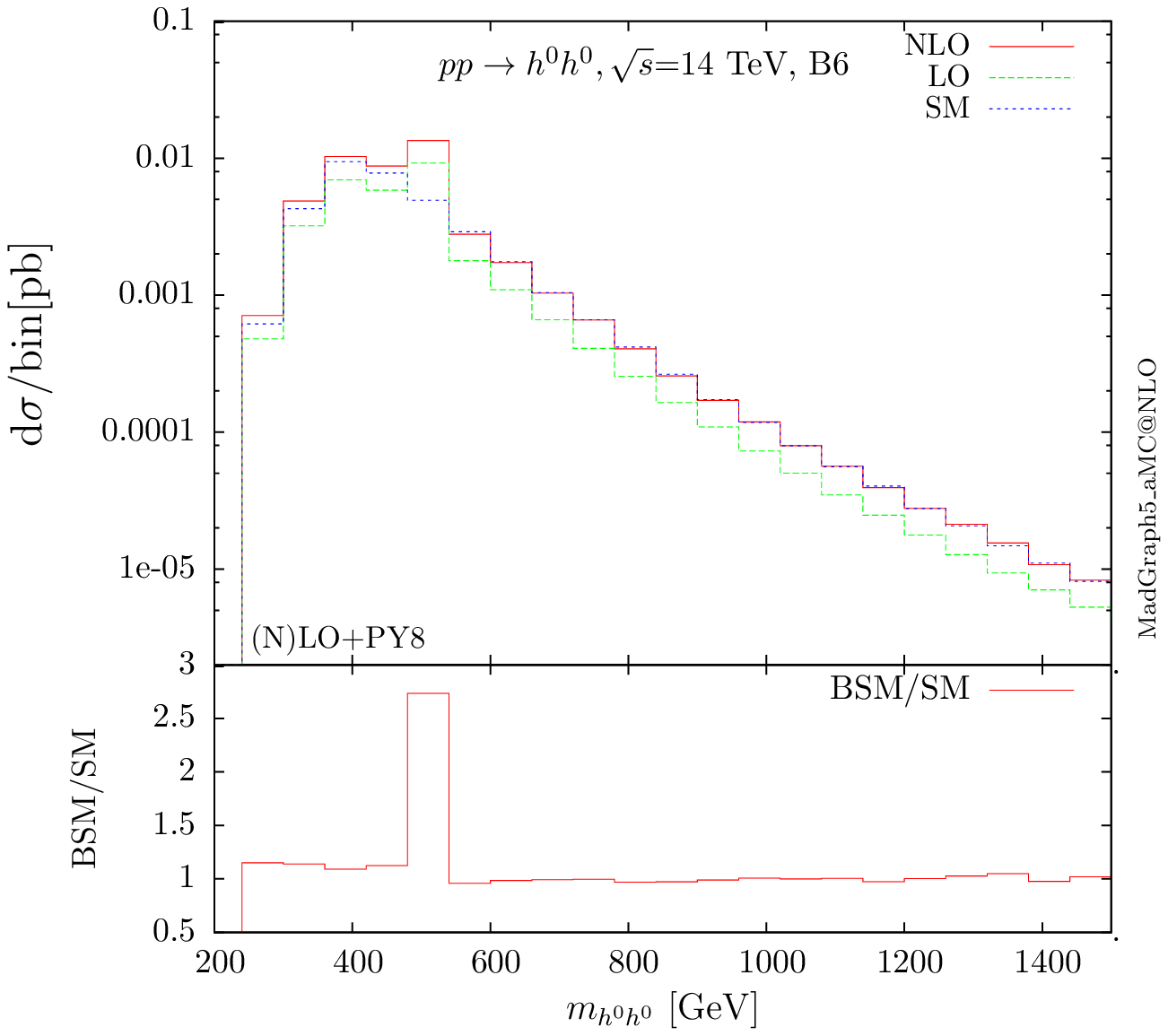}
\end{minipage}
\hspace{0.5cm}
 \begin{minipage}[b]{0.5\linewidth}
 \centering
 \includegraphics[trim=1.5cm 0 0 0,scale=0.56]{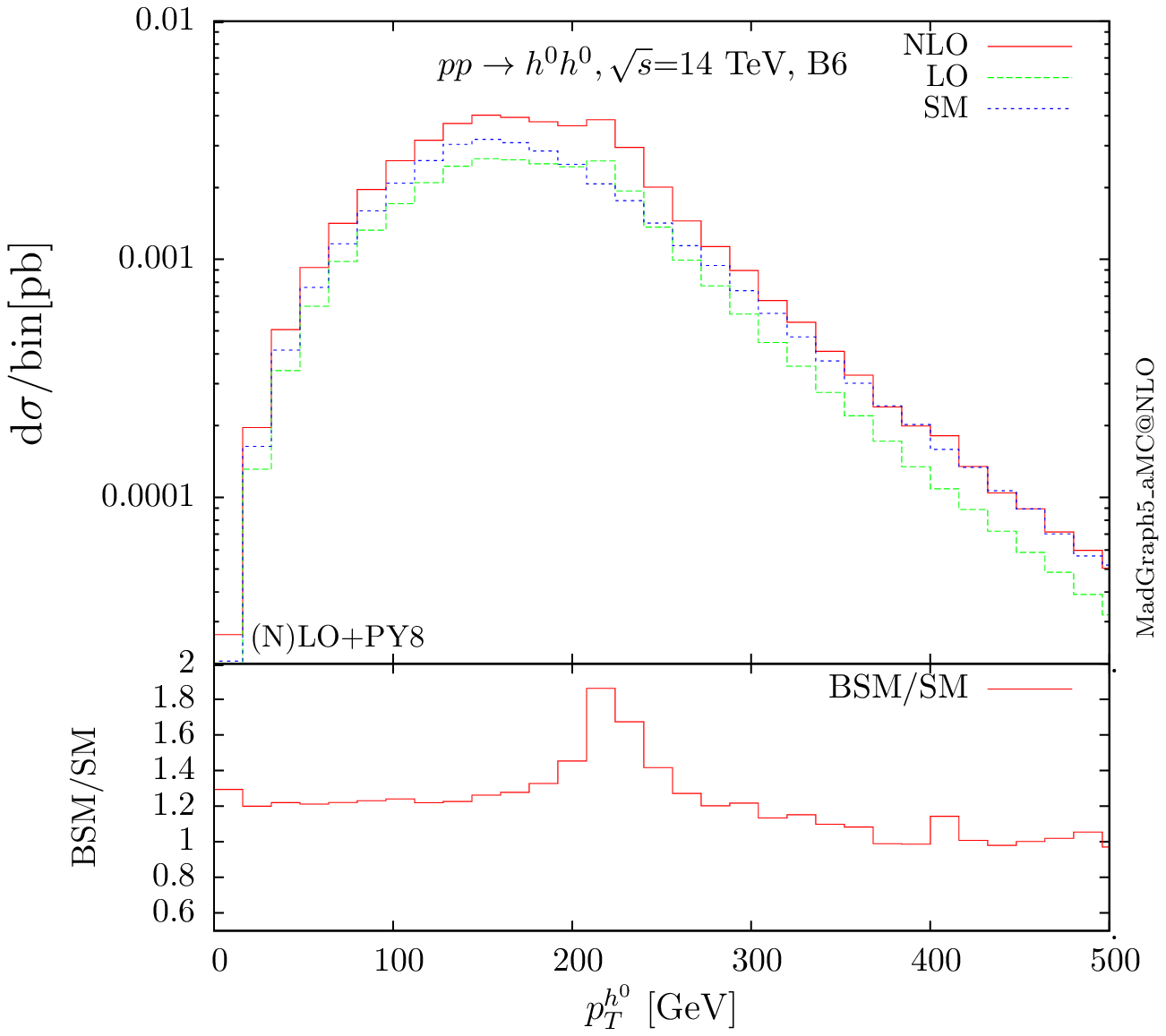}
 \end{minipage}
 \caption{\label{fig:b6} Light Higgs pair differential rates as a function 
of a) the di--Higgs invariant mass 
$m_{\hzero\hzero}$ (left panels); 
and b) the hardest Higgs transverse momentum $p^{\hzero}_T$ (right panels), in the same setup as for Figure~4. The 2HDM parameters are fixed as in benchmark B6.} 
\end{figure}

\begin{figure}[tb!]
 \begin{minipage}[b]{0.5\linewidth}
\centering
\includegraphics[trim=1cm 0 0 0,scale=0.56]{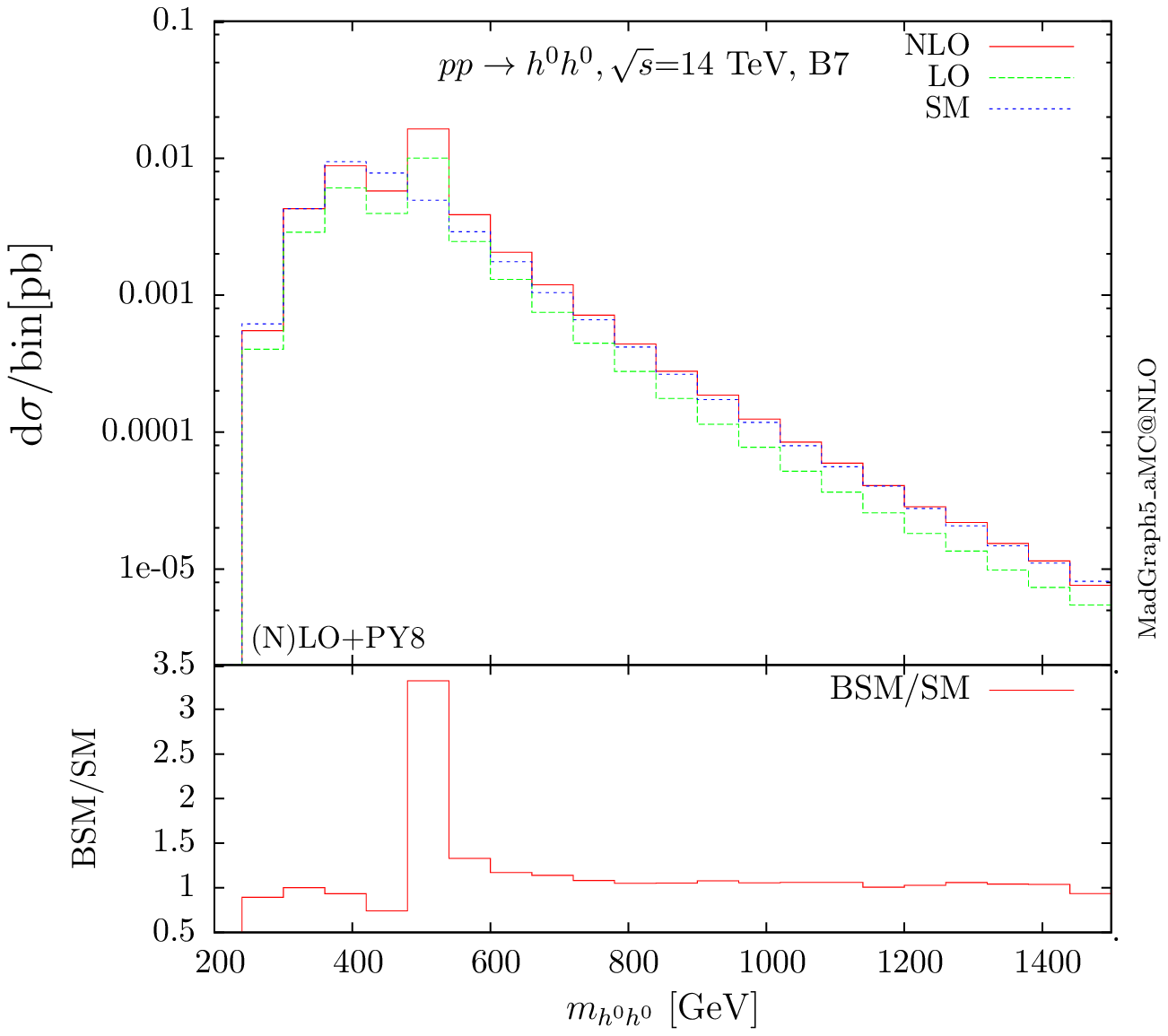}
\end{minipage}
\hspace{0.5cm}
 \begin{minipage}[b]{0.5\linewidth}
 \centering
 \includegraphics[trim=1.5cm 0 0 0,scale=0.56]{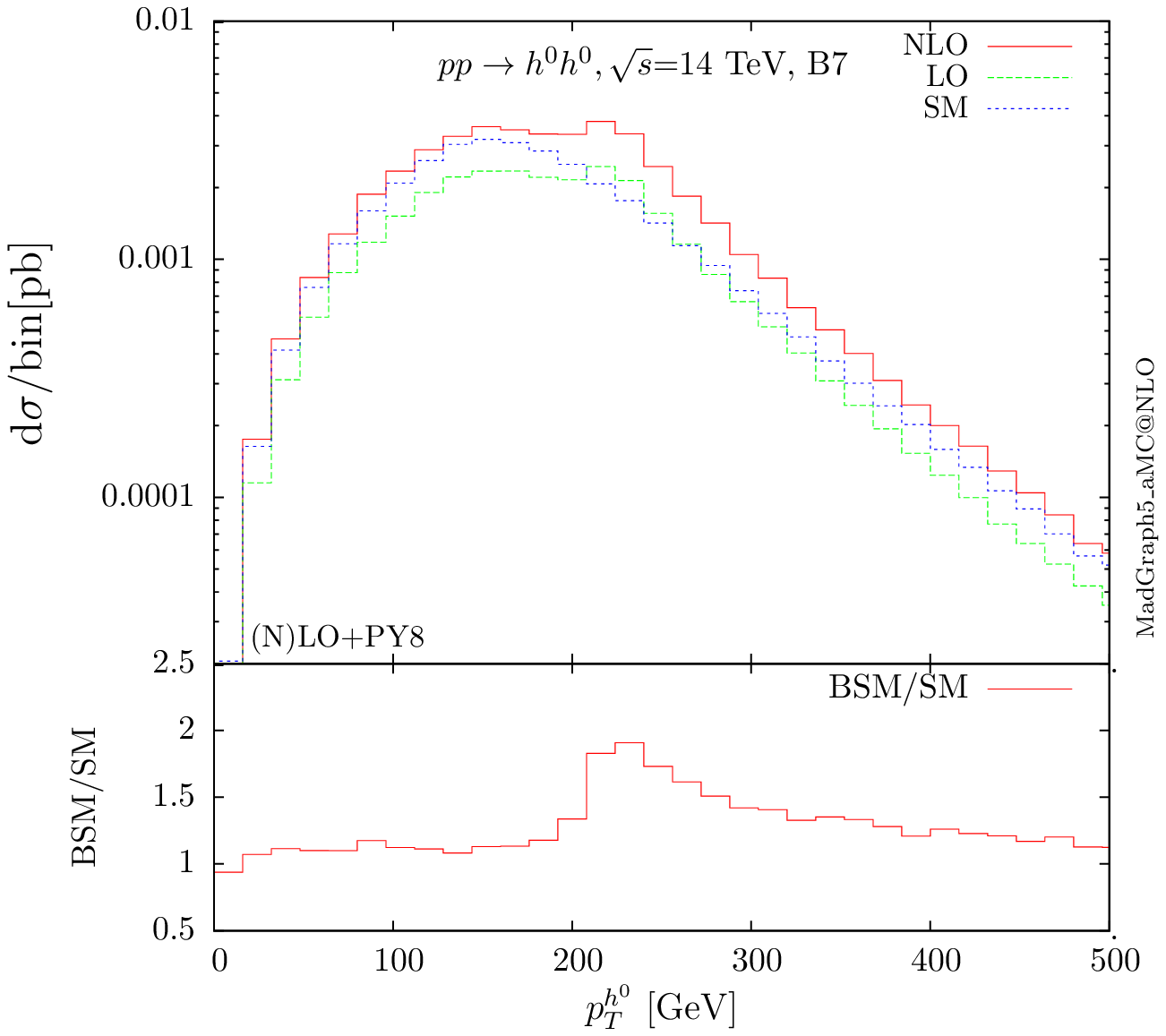}
 \end{minipage}
 \caption{\label{fig:b7} Light Higgs pair differential rates as a function 
of a) the di--Higgs invariant mass 
$m_{\hzero\hzero}$ (left panels); 
and b) the hardest Higgs transverse momentum $p^{\hzero}_T$ (right panels), in the same setup as for Figure~4. 
The 2HDM parameters are fixed as in benchmark B7.} 
\end{figure}

\smallskip{}
Instead, a richer landscape opens up
as we move on to the kinematical distributions.
These are particularly helpful to track down the model--specific
effects which are otherwise smeared once we integrate over the whole phase space. 
For the remainder of this section we shall focus
on the light neutral \CP--even Higgs pair channel $\hzero\hzero$
and study representative di--Higgs distributions for the LHC at 14 TeV.
Our results are displayed
in Figures~\ref{fig:b1}--\ref{fig:b7}, in which we show the light Higgs pair rates
as a function of the di--Higgs invariant mass $m_{\hzero\hzero}$ (left panels)
and the hardest Higgs transverse momentum $p_T^{\hzero}$ (right panels).
We concentrate on these two for the sake of conciseness, even though any alternative 
distribution can be obtained at will, as our setup is fully differential.
All histograms are shown at both LO+PS and NLO+PS accuracy, where the NLO results
follow from the ``loop--improved'' approach.
The SM prediction (also at NLO+PS) is overlayed for comparison. In the lower subpanels
we display the bin--by--bin ratio of the 2HDM NLO+PS results over the SM values.

\medskip{}
We begin by pointing out
a number of features common to all 2HDM scenarios. 
Let us first recall that, unlike the SM,
there are two types
of triangle topologies that contribute in the 2HDM; one of them is linked to the s--channel exchange
of the light Higgs boson $\hzero$, while the other one proceeds through the heavy Higgs exchange $\Hzero$.
The first subprocess $gg \to \hzero^* \to \hzero\hzero$
follows the shape of the 
virtual Higgs boson propagator and would peak around $\hat{s} \simeq m_{\hzero}^2$.
At larger invariant masses, 
the light Higgs propagator is probed off--shell, 
which means that for these
kinematical configurations, the light Higgs 
triangles become subleading.
The virtual heavy Higgs counterpart
$gg \to \Hzero^* \to \hzero\hzero$ 
becomes relevant at larger invariant mass values $m_{\hzero\hzero}\sim m_{\Hzero}$ and
may either add to, or partially cancel, the light Higgs triangle amplitudes in this region --
depending on the relative signs of the different Yukawa and trilinear Higgs self--couplings.
If $m_{\Hzero} > 2m_{\hzero}$, the heavy Higgs is produced
on--shell and its resonant peak takes over. 
As alluded to above, these resonant situations 
overshadow all other possible new physics effects. In particular,
neither the total nor the differential rates are sensitive  
to  a possible enhancement or suppression of the trilinear Higgs self--couplings. 

\medskip{}
The interplay between the triangles and box topologies,
which have different phase space dependence, 
generates a variety of model--specific signatures which  
are reflected in the histograms. For instance, 
a modified trilinear coupling $g_{\hzero\hzero\hzero}$ 
will mostly reveal itself at low invariant masses. Instead, the boxes
are unresponsive to variations in the triple Higgs self--interactions.
Shifted Yukawa couplings will typically become more apparent 
in the larger $m_{\hzero\hzero}$ and $p^{\hzero}_T$ regions,  
through their influence on the box contributions. 
The latter topologies have a slower decrease with 
$m_{\hzero\hzero}$ and $p^{\hzero}_T$ compared to the triangles,
and hence
dominate in the hard Higgs tails.  
Independent
changes in the top and bottom--quark Yukawas, which are possible
for type--II models, can in addition modify
the interference between the top and the bottom
loops.

\medskip{}
We can see from all histograms that, barring the 
cases with low--mass $\Hzero$ resonances,
the majority of the di--Higgs events
concentrates on the invariant masses well above the Higgs pair
threshold 
$ \hat{s} \gtrsim 4m_{\hzero}^2$.  This is after
all the reason for the breakdown
of the infinite top--mass effective theory (HEFT),
which is meant to hold for $\hat{s} \ll m_t^2$.
The HEFT fails to correctly reproduce the exact distributions
not only at
large invariant masses but also for moderate values.
The transverse momentum distributions are problematic for the same reason
\cite{Baur:2002rb,Gillioz:2012se}. 

\medskip{} We also notice that the QCD NLO effects,
while quantitatively important rate--wise, have very little
effect on the distribution shapes. Notice that
both the LO and the NLO histograms in all figures
vary in parallel, which implies a fairly constant $K$--factor. This is certainly not unexpected, 
in view of the structure of the NLO QCD corrections, as we have described in Section~\ref{sec:nlo}.
Neither the exchange of virtual gluons between the incoming partons nor the light parton radiation off the initial--state colored legs can significantly influence the leading kinematical features appearing 
in the $m_{\hzero\hzero}$ and $p_T^{\hzero}$ distributions, 
which rely fundamentally on ii) the relative sizes and signs of the heavy--quark Yukawa and the triple Higgs self--couplings;
and iii) the potential enhancement due to a resonant
(or close--to--resonant) intermediate heavy Higgs exchange -- all these mechanisms being already present at the LO.


\medskip{}
Let us now move on to the different model--specific features
which can be appreciated in the plots. 
We begin in Figure~\ref{fig:b1} by showing the results for
benchmark B1. The on--shell heavy Higgs contribution $gg \to \Hzero \to \hzero\hzero$
is evident in the resonant peaks for both the di--Higgs invariant mass
($m_{\hzero\hzero}\sim M_{\Hzero}=300$GeV) and the 
hardest Higgs transverse momentum distributions 
 ($p_T^{\hzero}\sim\sqrt{(m_{\Hzero}/2)^2-m_{\hzero}^2}$), which of course overwhelm
 the SM expectations in the low $m_{\hzero\hzero}$ and $p^{\hzero}_T$ regions. The dip
 in the signal right after the resonant peak is due to the interference between the boxes and the heavy Higgs--mediated triangles.
The small deviations from the SM away from the resonant peak are caused by the modified trilinear and Yukawa couplings. 
As expected, at large invariant masses $m_{\hzero\hzero} \gtrsim 600$ GeV the cross section is   
dominated by the box diagrams. Thereby we explain the 2HDM/SM ratio through the 
rescaled top Yukawa $(g_{\hzero tt}/g^{\text{SM}}_{Htt})^4 \sim 0.85$.

A qualitatively similar situation is encountered for benchmark B2, as shown in Figure~\ref{fig:b2}.
Again, the heavy Higgs resonant peak is manifest for $m_{\hzero\hzero} \simeq 700$ GeV and $p_T^{\hzero} \simeq 330$ GeV.
Given that the heavy Higgs mass is now larger, 
its on--shell single production via $gg \to \Hzero$ is suppressed by phase
space and by the lower gluon luminosity. This accounts for
the smaller rates 
with respect to the  B1 scenario
discussed above.
Close to the light Higgs pair threshold $m_{\hzero\hzero} \simeq 2m_{\hzero}$,
we find an
enhanced differential rate with respect to the SM. The interplay between several
effects is responsible for this behavior. 
On the one hand, 
the heavy Higgs--mediated triangles are 
small in these bins, while the light Higgs--mediated ones dominate. 
On the other hand, these leading triangle topologies are pulled down by the
$\mathcal{O}(40)\%$ reduction of the trilinear Higgs self--coupling $g_{\hzero\hzero\hzero}$.
The net result is a reduced (destructive) interference with the box contributions. 
The sharp dip
immediately after the resonant peak is once more due to the 
interference between the heavy Higgs triangles and the boxes -- which 
again leads to a partial cancellation in this case, because the $g_{\Hzero tt}$ and 
$g_{\Hzero \hzero\hzero}$ couplings 
have the same (negative) sign. 
In the large $m_{\hzero\hzero}$ tail, instead, the differential rates
mostly depend on the box contributions, and hence lie
roughly $\sim$ 20\% 
below the SM yields. 

\medskip{}
In contrast to the previous cases, the differential distributions in scenario B3
(cf. Figure~\ref{fig:b3}) barely depart from the 
SM. The reason is twofold: i)  
the absence of the on--shell heavy Higgs contribution;
ii) the SM--like pattern of all light Higgs couplings 
in the limit $\xi = \cos(\beta-\alpha) \to 0$.
In this case, 
also the
trilinear coupling $g_{\Hzero\hzero\hzero}$ is extremely suppressed,
so that the heavy Higgs--mediated process $gg \to \Hzero \to \hzero\hzero$ barely contributes.
The flat 2HDM/SM cross--section ratio is a further indication that  
no distinctive 2HDM imprints arise in any region of the phase space.
The results for benchmark B5, which we do not
show explicitly, are also featureless.  
The SM--like profile in the latter case also results from
the fermiophobic nature of the heavy Higgs boson,
which cannot couple to the quarks  
and hence has no influence on the light di--Higgs production. 

\medskip{}
A variety of non--standard imprints can be appreciated
in Figure~\ref{fig:b4} for benchmark B4. 
These genuine 2HDM effects have in this case
a non--resonant origin and can be
better interpreted by analysing
the individual components from the different topologies. The latter 
are shown separately in Figure~\ref{fig:mhhb4}.
One first observation is the increased rates right next to the di--Higgs threshold $m_{hh} \simeq 2m_{\hzero}$.
These are primarily caused
by the additional heavy Higgs contribution, and are particularly
strong in this case not only because of the low heavy Higgs mass $m_{\Hzero} = 200$ GeV, 
but also due to the unsuppressed trilinear coupling $g_{\Hzero\hzero\hzero}$ -- in contrast
to scenario B3 discussed above.
In fact, it turns out that in this region (viz. 
the first bin of Figure~\ref{fig:mhhb4}), the
box and the heavy Higgs triangle amplitudes
have comparable sizes but opposite signs. Given this
effective cancellation, the net result comes from the (enhanced) light Higgs triangles only. 
The sharp dip in the $m_{\hzero\hzero} \simeq 350$ GeV bin is related again to an accidental and nearly exact
cancellation between the three contributions in the game. In this
case, one can check that the two triangle--mediated contributions are largely cancelling the box--mediated ones.  
In the intermediate bins $m_{\hzero\hzero} \simeq 400 - 700$ GeV, 
the di--Higgs rates lie below the SM expectations due to the additional interference between the boxes
and the heavy Higgs--mediated triangles. 
This partial cancellation is quite strong in this case  
as both the top Yukawa $g^{\hzero}_{t}$ and the trilinear coupling $g_{\Hzero\hzero\hzero}$ are enhanced. 
This effect is also manifest in the transverse momentum distributions in the region
of 100 - 200~GeV (cf. right panels of Figure~\ref{fig:b4}). 
Finally, at large di--Higgs invariant masses and in the 
boosted Higgs tails (viz. $m_{\hzero\hzero} \gtrsim 800$ GeV
and $p_T \gtrsim 400$ GeV) the rates are dominated by the box
contributions. Correspondingly, the $\mathcal{O}(10\%)$--enhanced
Yukawa coupling accounts for 
the increased 
2HDM rates in these tails, which exceed the SM results by up to
$\mathcal{O}(50)\%$, as expected from the overall rescaling $(g_{\hzero tt}/g^{\text{SM}}_{Htt})^4 \simeq 1.5$. 

\medskip{}
Last but not least, in Figures~\ref{fig:b6} and \ref{fig:b7} we study the impact
of a strongly modified bottom--quark Yukawa, as realized
by benchmarks B6 and B7. While the two scenarios share a number
of similar features, their differences highlight some remarkable properties of
the 2HDM. In both cases we observe the expected on--shell peak
from the heavy Higgs cascade decay, although here
it is softer 
as compared to the previous resonant scenarios. 
This milder effect
is visible not only in the distributions but also rate--wise and
can be explained by the strongly suppressed top--quark Yukawa coupling
to the heavy Higgs $g_{\Hzero tt}$ (cf. Table~\ref{tab:couplings}).
One further common trait to both benchmarks is the asymptotic 
behavior at large di--Higgs invariant masses and in the boosted Higgs tails, where we obtain
results very close to the SM predictions.
All that said, we can also appreciate some relevant differences.
On the one hand, the $\mathcal{O}(20)$\% enhanced bottom Yukawa
in B6 reinforces the destructive interference between the top and bottom--mediated
triangles. In addition, the amplitude of the individual triangle diagrams
is pulled down further by the (slightly suppressed) trilinear self--coupling $g_{\hzero\hzero\hzero}$.
Both effects cooperate to reduce the interference between the
triangle and the box topologies in the lowest $m_{\hzero\hzero}$ and $p^{\hzero}_T$ bins,
in such a way that we obtain rates slightly above the SM expectations.

\medskip{}
A similar mechanism operates in
the B7 scenario, albeit in the opposite direction. 
In this case, the sign--flipped
bottom--quark Yukawa causes the 
top and bottom--mediated triangles to interfere constructively.
As a result, we are left with slightly enhanced triangle amplitudes,
which thus reinforce the interference with the boxes.
In agreement with this fact, 
the predicted number of events for $m_{\hzero\hzero}\lesssim m_{\Hzero}$ falls
slightly below the SM expectations. 
Unlike the previous resonant scenarios, in this case
we find no dip right after the resonance peak.
Instead, we obtain slightly enhanced rates for $m_{\hzero\hzero} \gtrsim m_{\Hzero}$ because the
fact that $(g_{\Hzero\hzero\hzero}) (g_{\Hzero tt}) < 0 $ (cf. Table~\ref{tab:couplings})
leads to a constructive interference with the boxes.  
A dip does appear instead right below $m_{\hzero\hzero} \lesssim m_{\Hzero}$ due
to the additional negative sign from the heavy Higgs propagator $1/(s-m^2_{\Hzero})$.

\medskip{}
At this point, let us mention that we are aware of
possible caveats in matching the NLO prediction to  
the parton shower in the case of enhanced bottom Yukawas.
These have been discussed in the context of single Higgs production~\cite{Bagnaschi:2011tu,Harlander:2013qxa}
and are related to the heavy--quark mass dependence of the Higgs transverse momentum
distributions, which is not known exactly beyond the LO.
These issues mostly concern the low Higgs $p_T$ region and become more
relevant if the bottom--quark Yukawa increases. 
This is indeed the reason why we do not expect them to matter in our case, as in all of the scenarios 
that we have explored only the bottom--quark
coupling to the heavy Higgs boson $g_{\Hzero bb}$ is significantly enhanced. 
In view of this fact, and since the bottom--mediated effects to the 
$gg$--induced heavy Higgs production are very small, 
we do not find it necessary to adjust our setup to include a more dedicated treatment.

\section{Conclusions}
\label{sec:conclusions}

The production of Higgs boson pairs at hadron colliders
provides a direct handle on the trilinear Higgs self--coupling.
Its direct experimental measurement plays a paramount role 
in the reconstruction of the Higgs potential and represents
a key step towards unraveling the fundamental structure of electroweak symmetry breaking.
While the profile of the $\sim 126$ GeV resonance discovered at the LHC
largely agrees with a SM Higgs profile, there is still experimental room, and also strong 
theoretical motivation, for a non--minimal Higgs sector.
Accurate predictions for Higgs production
cross sections and distributions, including
reliable estimates on the theoretical uncertainties, are thus necessary
to identify and characterize eventual deviations from the SM.
The 2HDM constitutes an exemplary framework 
in which to describe the possible signatures of an extended Higgs sector.
The model--specific imprints on the Higgs pair production phenomenology at the LHC
can have a resonant or non--resonant
origin and may be categorized as follows: 
i) direct production of additional Higgs boson pairs; ii)  
virtual or resonant heavy Higgs--mediated contribution to the production of light Higgs pairs; 
iii) modified Higgs couplings. 

\medskip{}
In this paper we have examined the production of Higgs boson pairs
via gluon fusion in the 2HDM.
We have provided
predictions for the 14 TeV LHC at NLO accuracy in QCD, matched to the {\sc Pythia8} parton
shower. We have considered all seven possible final--state Higgs pair combinations
in the model and explored representative up--to--date 2HDM benchmarks.
The Higgs bosons in the final state have been kept stable in our simulated
event samples.
Our calculations have been based on the publicly available tool MadGraph5$\_${ \sc aMC@NLO}
together with the {\sc 2HDM@NLO} model implementation obtained with the {\sc NLOCT} package.
The NLO QCD corrections to the (loop--induced) gluon fusion 
channels have been handled via a reweighting procedure which
includes the exact one--loop matrix elements. Dedicated codes for this computation are
available in \cite{web}.

\medskip{}
We have reported large QCD corrections, reflected in the sizable
$K$--factors in the ballpark of $K\sim 1.5-1.7$ (for the $gg$--initiated
channels) and $K \sim 1.2-1.3$ (for the $q\bar{q}$--initiated ones).
These QCD effects are dominated by the initial--state light--parton
radiation. They remain almost constant 
over the phase space and barely depend on the model parameters. 
Once they are taken into account, 
the theoretical uncertainties on the Higgs pair rate predictions are significantly reduced.

\medskip{}
We have examined a variety of characteristic 2HDM features
and evaluated their effect on the total Higgs pair rates and kinematical distributions.
The underlying model structure influences the light di--Higgs production 
in different ways i) the virtual (real) heavy Higgs--mediated
contribution $gg \to \Hzero^{*}(\Hzero) \to \hzero\hzero$,
which may enhance the total rates by up to roughly 2 orders of magnitude above the
SM expectations; ii)
the diverse possible combinations of enhanced, suppressed and/or sign--flipped Higgs couplings,
which lead to increased or reduced rates, particularly
apparent in the differential distributions.

\medskip{} 
One further step in this direction is to combine our current
results with the corresponding 
Higgs branching ratios, and to evaluate 
the signal--over--background significances for the respective decay modes.
Identifying the channels with the best prospects of being measured  
requires dedicated studies which should involve detector effects and selection cuts to
optimize the signal extraction over the backgrounds. 
Complementary information can be obtained 
by including the additional double Higgs mechanisms 
besides the dominant gluon fusion mode, namely 
vector boson fusion;
associated production with gauge bosons; and Higgs radiation off
heavy quarks. The ultimate goal is to identify the most promising experimental opportunities
for these extended Higgs sector searches at the LHC.

\section*{Acknowledgements}
The authors are indebted to C\'eline Degrande, Fabio Maltoni and Marco Zaro for enlightening discussions, technical assistance
and constant encouragement in the course of this project.
This work has been supported in part by the Research Executive Agency (REA) of the 
European Union under the Grant Agreement number PITN-GA-2012-315877 (MCNet) and by the National Fund for Scientific Research (F.R.S.-FNRS Belgium) under a FRIA grant. 
DLV acknowledges the support of the   
F.R.S.-FNRS ``Fonds de la Recherche Scientifique'' (Belgium).

\bibliographystyle{JHEP}
\bibliography{hhbib}
\end{document}